%% file: main.tex
\documentclass[fleqn,usenatbib]{mnras}
\usepackage[T1]{fontenc}
\usepackage{ae,aecompl}
\usepackage{epsfig,color}
\usepackage{amsmath}
\usepackage{graphicx}
\usepackage{amssymb}

\newcommand{\rmunit}{\mbox{$\rm \,rad \,m^{-2}$}}

\title[The C-Band All-Sky Survey: Design]{The C-Band All-Sky Survey (C-BASS): Design and capabilities}
\author[M. E. Jones et al.]{Michael~E.~Jones,$^{1}$\thanks{E-mail: \url{mike.jones@physics.ox.ac.uk}}  Angela~C.~Taylor,$\!^{1}$ Moumita Aich,$\!^2$ C.~J.~Copley,$\!^{1,3}$ 
\newauthor H. Cynthia Chiang,$\!^2$ $\,$R.~J.~Davis,$\!^{4}$\thanks{Deceased} C.~Dickinson,$\!^{4}$ R.~D.~P.~Grumitt,$\!^{1}$ Yaser~Hafez,$\!^{5}$  
\newauthor Heiko M. Heilgendorff,$\!^2$ C.~M.~Holler,$\!^{1,6}$  M.~O.~Irfan,$\!^{4,7}$ 
 Luke R.~P.~Jew,$\!^{1}$ J.~J.~John,$\!^{1}$ \newauthor  J.~Jonas,$\!^{3}$ O.~G.~King,$\!^{1,8}$ J.~P.~Leahy,$\!^{4}$  J.~Leech,$\!^{1}$ E. M. Leitch,$\!^{8}$ S.~J.~C.~Muchovej,$\!^{8}$  \newauthor T.~J.~Pearson,$\!^{8}$ M. W. Peel,$\!^{4,9}$ A.~C.~S.~Readhead,$\!^{8}$ Jonathan Sievers,$\!^2$ M.~A.~Stevenson$^{8}$ \newauthor and J.~Zuntz$^{1,4,10}$ 
 \\ 
$^{1}$Sub-department of Astrophysics, University of Oxford, Denys Wilkinson Building, Keble Road, Oxford OX1 3RH, UK \\
$^{2}$Astrophysics \& Cosmology Research Unit, School of Mathematics, Statistics \& Computer Science, University of KwaZulu-Natal, \\Westville Campus, Private Bag X54001, Durban 4000, South Africa \\
$^{3}$Department of Physics and Electronics, Rhodes University, Drostdy Road, Grahamstown, 6139, South Africa \\
$^{4}$Jodrell Bank Centre for Astrophysics, School of Physics and Astronomy, The University of Manchester, Manchester, M13 9PL, UK \\
$^{5}$King Abdulaziz City for Science and Technology, Saudi Arabia\\
$^{6}$Munich University of Applied Sciences, Lothstra{\ss}e 34, Munich 80335, Germany \\
$^{7}$Laboratoire CosmoStat, AIM, UMR CEA-CNRS-Paris 7 Irfu, SAp/SEDI, Service d'Astrophysique, CEA Saclay, France\\
$^{8}$California Institute of Technology, Pasadena, CA 91125, USA \\
$^{9}$Departamento de F\'{i}sica Matematica, Instituto de F\'{i}sica, Universidade de S\~{a}o Paulo, Rua do Mat\~{a}o 1371, S\~{a}o Paulo, Brazil\\
$^{10}$Institute for Astronomy, University of Edinburgh, Edinburgh, EH9 3HJ, UK
}

\pubyear{2018}
\date{Accepted 2018 July 19. Received 2018 July 17; in original form 2018 May 11}
\begin{document}
\label{firstpage}
\pagerange{\pageref{firstpage}--\pageref{lastpage}}
\maketitle

\setcounter{tocdepth}{3}


\begin{abstract}

  The C-Band All-Sky Survey (C-BASS) is an all-sky full-polarization
  survey at a frequency of 5\,GHz, designed to provide complementary
  data to the all-sky surveys of {\it WMAP} and {\it Planck}, and future CMB
  $B$-mode polarization imaging surveys. The observing frequency has
  been chosen to provide a signal that is dominated by Galactic
  synchrotron emission, but suffers little from Faraday rotation, so
  that the measured polarization directions provide a good template
  for higher frequency observations, and carry direct information
  about the Galactic magnetic field. Telescopes in both northern and
  southern hemispheres with matched optical performance are used to
  provide all-sky coverage from a ground-based experiment. A
  continuous-comparison radiometer and a correlation polarimeter on each
  telescope provide stable imaging properties such that all angular
  scales from the instrument resolution of 45\,arcmin up to full sky
  are accurately measured. The northern instrument has completed its
  survey and the southern instrument has started observing. We expect
  that C-BASS data will significantly improve the component separation
  analysis of {\it Planck} and other CMB data, and will provide important constraints on
  the properties of anomalous Galactic dust and the Galactic magnetic
  field.

\end{abstract}

\begin{keywords}
methods: data analysis -- radio continuum: general -- techniques: image processing -- diffuse radiation -- cosmic microwave background
\end{keywords}


\input{intro}
\input{surveys}
\input{foregrounds}

\input{requirements}
\input{design}
\input{analysis}

\input{impact}

\section{Conclusions}
\label{sec:conclusions}

Low-frequency radio surveys are an essential component of a CMB
foreground removal strategy, providing constraints on the synchrotron,
free-free and AME components of Galactic emission. However, all-sky
surveys to date below 20\,GHz have been of limited use due to map
artefacts and calibration problems. The C-Band All-Sky Survey will
provide accurate and well-calibrated maps of the whole sky in Stokes
$I$, $Q$ and $U$ at 5\,GHz, with additional frequency resolution in the
southern part of the survey. This will allow a major improvement in
the accuracy of foreground separation for CMB intensity and
polarization measurements. The data will also be used to study diffuse Galactic emission, such as measuring the synchrotron spectral index, constraining foreground models for studying AME at higher frequencies, and constraining models of the Galactic magnetic field. 

The northern survey is now complete, with the telescope having been decommissioned in April 2015. Data reduction and analysis for the northern data are ongoing, and full results will be presented in forthcoming papers. Preliminary maps of the northern sky have been presented by \citet{moriond2018}. At the time of writing, observations were still being made for the southern survey. 

The C-BASS frequency at 5~GHz is the ideal balance between being sufficiently low to give good sensitivity to synchrotron radiation, with its steeply falling spectrum, and sufficiently high to avoid the worst effects of depolarization and Faraday rotation. Higher sensitivity observations at frequencies above C-BASS but below the space microwave band would of course give even better constraints on the synchrotron spectrum. C-BASS has been designed to give a clean beam with relatively high main-beam efficiency, well understood sidelobe structure, and minimal far-out and cross-polarization sidelobes. This allows accurate calibration and gives a well-understood effective temperature scale. The inclusion of C-BASS data in component separation analyses will break degeneracies in both intensity and polarization measurements, allowing more accurate estimation of foregrounds and hence of the CMB component. This additional accuracy will be crucial for future $B$-mode detections.

\section*{Acknowledgments}

The C-BASS project is a collaboration between Oxford and Manchester Universities in the U.K., the California Institute of Technology in the U.S., Rhodes University, UKZN and the South African Radio Astronomy Observatory in South Africa, and the King Abdulaziz City for Science and Technology (KACST) in Saudi Arabia.
The work at Oxford was supported by funding from STFC, the Royal Society and the University of Oxford. The work at the California Institute of Technology and Owens Valley Radio Observatory was supported by National Science Foundation (NSF) awards~AST-0607857, AST-1010024, AST-1212217, and AST-1616227, and by NASA award NNX15AF06G. The work at Manchester was supported by STFC and CD also acknowledges support from an ERC Starting (Consolidator) Grant (no.~307209). OGK acknowledges the support of a Dorothy Hodgkin Award in funding his studies while a student at Oxford, and the support of a W.M. Keck Institute for Space Studies Postdoctoral Fellowship at Caltech. CJC acknowledges the support of a Commonwealth Scholarship in funding his studies while a student at Oxford. MP acknowledges funding from a FAPESP Young Investigator fellowship, grants 2015/19936-1 and 2016/19425-0, S\~{a}o Paulo Research Foundation (FAPESP). HMH acknowledges the financial assistance of the South African SKA Project (SKA SA) (www.ska.ac.za) towards this research. We also thank Hans Kristian Eriksen and Ingunn Wehus for their assistance with producing Fig.~\ref{fig:frequency_spectra}. Finally, we thank the late Profs. Richard J. Davis and Rodney D. Davies, who were strong supporters of the C-BASS project from the beginning. \url{http://cbass.web.ox.ac.uk}

\bibliographystyle{mnras}
\bibliography{refs}
\bsp


\label{lastpage}

\end{document}

%% file: intro.tex
\section{Introduction}

In recent years great effort has been made to systematically survey
the whole sky from microwave to sub-millimetre wavelengths using the
\emph{WMAP} \citep{Bennett2013} and \emph{Planck} \citep{Planck2015_I}
spacecraft. These surveys have primarily been aimed at studying the
cosmic microwave background (CMB) radiation, and have yielded
cosmological information of unprecedented precision \citep{Hinshaw2013,Planck2015_I}. 

Since the first searches for anisotropies in the CMB, the danger
that foreground emission could masquerade as the sought-for
cosmological signal has been of great concern. Consequently, most CMB
experiments have involved observing at multiple frequencies. This was
first done to confirm the expected thermal spectrum of the
anisotropies \citep[e.g.,][]{Smoot1992}. In later experiments, cuts on
the sky were defined, in frequency, and in angular scale (multipole
range) where CMB fluctuations were known to dominate over foregrounds
\citep[e.g.,][]{Planck2015_XI}, so that only minor foreground
corrections were needed.

The practical limit to this strategy has now been reached with the
attempt to detect large-scale $B$-mode fluctuations in the CMB
polarization \citep{Zaldarriaga1997,KKS1997}, which would be convincing
evidence of the reality of inflation and would determine the
characteristic energy of the inflaton field.  A recent claimed
detection of inflationary $B$-modes from the BICEP2 experiment
\citep{BICEP2014}, in a region selected specifically for minimal
foreground emission,
has now been explained in terms of polarized thermal dust emission
\citep{Bicep-Planck}.  Evidently, in future we will need to model and
subtract foregrounds with high accuracy, to reveal CMB signals that are subdominant at all frequencies.

Early hopes that multifrequency analyses using the wealth of frequency
channels obtained by \emph{WMAP} and \emph{Planck} would allow
accurate foreground correction have been only partially fulfilled
\citep[e.g.,][]{Planck2015_X}.  Foreground emission has a minimum
brightness relative to the CMB at around 70\,GHz.  While \emph{Planck}
has mapped the dominant high-frequency component (thermal dust
emission) to high enough frequencies that the CMB fluctuations
themselves are negligible and the foreground is well detected all over
the sky, on the low frequency side the foregrounds remain subdominant
to the CMB fluctuations at high Galactic latitudes at the lowest
frequency observed from space, the \emph{WMAP} 23\,GHz channel.
Furthermore, the low-frequency foreground spectrum has proved
substantially more complicated than was expected when the frequency
coverage of these instruments were designed.  Originally, it was
believed to consist of free-free and synchrotron emission, but we now
know there is a third continuum component, termed anomalous microwave
emission (AME; \citealt{Leitch1997}). Moreover, the synchrotron component is spectrally more
complicated than anticipated (see Section\,\ref{sec:synchrotron}).
Consequently, in the narrow band (23--70\,GHz) where these three
mechanisms are detected by the CMB spacecraft, they cannot be reliably
disentangled \citep[e.g.,][]{Planck2015_XXV}.

For more reliable modelling, we need to extend the frequency coverage
to much lower frequencies, where the spectra of the three
low-frequency components should be easily distinguishable
\citep{Krachmalnicoff2015,Remazeilles2016}.  This will also give sky
maps where the low-frequency foregrounds are clearly detected in each
pixel.  These observations must be carried out from the ground,
because wavelengths much longer than 1\,cm are not practical for CMB
space missions, due to the large size of the feeds required and the
limited resolution available from the relatively small size of the primary
mirror.

In this paper we describe the design, specifications, and capabilities
for one such project: the C-Band All-Sky Survey
(C-BASS)\footnote{\url{http://cbass.web.ox.ac.uk}}, which
aims to map the entire sky in total intensity and polarization at
5\,GHz, at a resolution of $45$~arcmin. 5 GHz is
simultaneously the highest frequency at which the foreground
polarization will be clearly detected all across the sky, and the
lowest frequency at which the confusing effects of Faraday rotation
and depolarization can be robustly corrected.  The survey is being conducted
in two parts, a northern survey using a 6.1-metre telescope at the
Owens Valley Radio Observatory (OVRO) in California, and a southern
survey with a 7.6-metre telescope at Klerefontein in South
Africa. Although the telescopes are somewhat different in size, the
optics are designed to give the same beamsize with both instruments
\citep{Holler2011}. The instruments are designed to provide a
high-efficiency beam with low intrinsic cross-polarization, and to
have sufficient stability to produce maps not limited by systematic
effects.  The C-BASS
maps will enable new studies of the interstellar medium and magnetic
field in the Galaxy, and help to determine the origin of the
poorly-understood anomalous microwave emission (AME). They will be
used to model the polarized synchrotron emission from the Galaxy; this
model will be essential for removing foreground emission from the
cosmic microwave background polarization maps from \emph{WMAP}, {\it
  Planck}, and future CMB missions.

The remainder of this paper is organised as follows. Section
\ref{sec:surveys} summarises the existing large-area radio and
microwave surveys, and Section~\ref{sec:mechanisms} reviews the foreground
emission mechanisms that need to be measured and modelled, which motivated the design of C-BASS.
Section~\ref{sec:requirements} outlines the requirements for the
survey and instrument design necessary to achieve the
scientific goals of the project, and Section \ref{sec:design}
describes the instrument design adopted. In Section \ref{sec:data} we
describe how the raw data are calibrated and used to make the primary
science data products, which are maps of Stokes parameters.
Section~\ref{sec:impact} outlines the impact that C-BASS will have on
both CMB and Galactic science, and we summarise our conclusions in
Section~\ref{sec:conclusions}.

%% file: surveys.tex
\section{Large-area radio surveys}
\label{sec:surveys}

Table~\ref{tab:surveys} summarises the current state of large-area
surveys in the frequency range useful for modelling CMB foregrounds,
roughly 400 MHz to 1 THz \citep[for a discussion of radio surveys at
lower frequencies see][]{deOliveiraCosta2008}. The table only includes
surveys that cover at least $2\pi$~sr  and that have angular
resolutions of $\approx 1\degr$ or better. 

\setlength{\tabcolsep}{3pt}
\begin{table*}
\caption{Existing and on-going large-area radio surveys of intensity and polarization between 400\,MHz and 1\,THz,  and with angular resolutions $\lesssim 1\degr$.}
\label{tab:surveys}
\begin{tabular}{lllccrrll}
\hline
Survey / 	       & Frequency  & FWHM       & Declination 	       &Stokes$^a$    &\multicolumn{2}{c}{Sensitivity$^b$}  & Status$^c$	& Reference(s)\\   
Telescope              & [GHz]      & [arcmin]   & Coverage	       &  & noise & offsets	&  & \\
\hline
Haslam (various)       & 0.408	   & 51		& All-sky	 &$I$     & 1\,K   & 3\,K 	& 3 & \protect{\cite{Haslam1982}} \\ 
Dwingeloo	&0.82	&72		&$-7^{\circ}$ to $+85^{\circ}$  &$I$	&0.2\,K		&0.6\,K	&3	&\protect\cite{Berkhuijsen1972}  \\
CHIPASS (Parkes)       & 1.394     & 14.4       & $< +25\degr$ &$ I$    & 0.6\,mK  & 30\,mK        & 3 &  \protect\cite{Calabretta2014} \\
DRAO (26-m)$^d$  & 1.4	   & 36		& $> -29\degr$   &$QU$    & 12\,mK & 30\,mK	& 3 &  \protect\cite{Wolleben2006} \\
Villa Elisa$^d$  & 1.4 	   &35.4        & $< +10\degr$   &$IQU$   & 9\,mK  & 50\,mK	& 3 &  \protect\cite{Testori2008} \\ 
Stockert$^d$     & 1.42	   &35		& $> -30\degr$   &$I$     &  9\,mK & 50\,mK	& 3 & \protect\cite{Reich1986}\\ 
GMIMS-HB N             &1.28--1.75 & 30         & $> -30\degr$   & $IQU$  & 12\,mK & unknown    & 1 &  \protect\cite{Wolleben2010} \\
STAPS (Parkes)         &1.3--1.8   & 15         & $<  0\degr$   & $IQU$  & unknown & unknown  &  1 & Haverkorn (priv. comm.) \\
HartRAO	               & 2.326	   & 20		& $-$83\degr\ to $+$13\degr &$I-Q$     & 25\,mK &80\,mK	& 3 & \protect\cite{Jonas1998} \\ 
S-PASS (Parkes)        & 2.3       & \phantom{0}9         & $< 0\degr$     &$IQU$   & 0.1\,mK & unknown      & 1    & \protect \cite{Carretti2013} \\ 
GEM                    & 4.8--5.2  &  45        & $-$52\degr\ to $+$7\degr& $QU$ & 0.5\,mK         & unknown & 0 & \protect\cite{Barbosa2006,Tello2013}\\
C-BASS                 & 4.5--5.5  &  45        & All-sky               & $IQU$  & 0.1\,mK         & 1 mK  & 0 &  This paper\\
QUIJOTE                & 11--19,30,40    &$\approx 60$& $\gtrsim 0\degr$ &$[I]QU$ & $25\,\mu$K & unknown & 1 & \protect\cite{Genova-Santos2015a} \\      
{\it WMAP}             & 22.8--94  & 49--15     & All-sky	        &$IQU$   &$4\,\mu$K & 1\,$\mu$K & 3	& \protect\cite{Bennett2013} \\ 
{\it Planck} LFI       &28.4--70   & 32--13     & All-sky	        &$IQU$   &$3\,\mu$K & 1\,$\mu$K & 2	& \protect\cite{Planck2015_I} \\ 
{\it Planck} HFI       &100--353   & 10--5      & All-sky	        &$IQU$   & 0.2--0.5\,$\mu$K   & 1--5\,$\mu$K  & 2	& \protect\cite{Planck2015_I} \\ 
{\it Planck} HFI       &545, 857   & \phantom{0}5          & All-sky	        &$I$     & 0.4, 0.8\,$\mu$K   & 1\,$\mu$K  & 2 & \protect\cite{Planck2015_I} \\ 
CLASS                  & 38--217   & 90--18     & $-$68\degr\ to $+$22\degr& $QU$     & 0.4\,$\mu$K   & unknown & 0 & \protect\cite{Harrington2016}\\
\hline
\end{tabular} \\
$^a$ [I]QU denotes surveys where total intensity (Stokes I) is measured but with much larger systematic errors than
for the linear polarization (Stokes Q and U). I$-$Q denotes a single linear polarization.\\
$^b$ Approximate average total intensity sensitivity in Rayleigh-Jeans temperature after convolution to 1\degr\ FWHM resolution: 
``noise'' is local rms; ``offsets'' is global systematic uncertainty. \\
$^c$ Status 0: observations ongoing; 1: observations complete, reduction in progress; 2: preliminary results released;
3: Final data released. \\
$^d$ An all-sky 1.4\,GHz map in IQU has been assembled from the
Stockert, DRAO and Villa Elisa surveys  \protect\citep{Reich2004,Testori2008},
but full details of its construction have not been published, and it is
not clear if the currently-available version
is the final one.
\end{table*}

 The separation of foregrounds from CMB emission places strong demands on
 the accuracy of the sky maps, which must be absolutely calibrated to of
 order 1 per cent precision, and must accurately reproduce sky features on
 scales of tens of degrees. Far sidelobe responses to the bright
 Galactic plane, the Sun and Moon, and the ground around the telescope
 must be reduced to well below the high-latitude foreground
 intensity. Even for the {\it Planck} spacecraft, with its unblocked
 optical system designed to minimize far sidelobes, this
 could only be achieved by correcting the maps for sidelobe responses;
 even then some detectors had to be omitted due to excessive
 residual sidelobes, to achieve the best multi-frequency fit
 \mbox{\citep{Planck2015_X}}.  

 The ground-based radio surveys published to date were never intended
 to reach this level of accuracy, and typically suffer from
 unquantified sidelobe responses \citep[see e.g.,][]{Du2016} and scan-synchronous artefacts in the
 maps, which limit the accuracy and fidelity of the images. For
 example, \citet{Calabretta2014} show a difference map between the
 1.4\,GHz Stockert/Villa Elisa and CHIPASS surveys, which reveals
 obvious scan-synchronous residuals. These features significantly
 degrade the recovered component maps if these surveys are included in
 component separation analysis, and in practice they do not add
 usefully to the analysis.  The most useful all-sky low-frequency survey for intensity measurements is
 the 408\,MHz survey of \citet{Haslam1982}. Although it also contains
 artefacts, there have been a number of attempts to remove the residual
 striping in this map, most recently and successfully by
 \mbox{\citet{Remazeilles2015a}}.  In practice this is the only ground-based
 survey that has proved useful in CMB component separation, thanks to a
 relatively clean beam, the high sky brightness which reduces the
 relative impact of ground pick-up, and to the long frequency lever arm
 to the space microwave band,\footnote{ By `microwave' we mean
   frequencies of 3--300\,GHz, while `space microwave' is the part of
   this band used by space survey missions, roughly 20--300\,GHz.}
 which reduces the impact of map errors on derived spectral indices.

 In polarization, the Villa Elisa and DRAO surveys at 1.4\,GHz are the
 only large-area surveys to have been fully published; but in any case
 at frequencies of a few GHz there is significant depolarization and
 polarization angle rotation due to Faraday rotation, which
 substantially complicates multi-frequency modelling of the sky
 polarization. We can estimate the size of the effect from the
 catalogue of Faraday rotation measures, RM, of extragalactic sources
 by \citet{Taylor2009}: at $|b| > 30\degr$ the rms rotation measure is
 $\sigma_{\rm RM} \approx 28$\rmunit, while at lower latitudes $\sigma_{\rm RM}
 \approx 85\rmunit$. We are primarily interested in the diffuse
 interstellar polarization, for which emission and Faraday rotation are
 mixed along the line of sight, giving RMs roughly half the
 extragalactic values, so the typical rotations at high (low) latitudes
 are 37\degr (112\degr) at 1.4\,GHz, 14\degr (42\degr) at 2.3\,GHz, and
 3\degr (9\degr) at 5\,GHz. Strong depolarization is likely to set in
 when rotations exceed about a radian, and indeed the sky polarization
 at $|b| < 30\degr$ towards the inner Galaxy is largely suppressed in
 the 1.4\,GHz surveys. These numbers illustrate one of our main motives
 for choosing to observe at 5\,GHz, but they also show that to
 accurately model the polarization in the space microwave band
 we will have to correct for the residual (few degrees  at most) Faraday rotation at 5\,GHz.

 Fortunately, two new surveys should yield the required RM data. The
 Global Magneto-Ionic Medium Survey (GMIMS) is an ambitious project to
 map the entire sky with continuous frequency coverage in the range
 0.3--1.8\,GHz, to allow high-resolution Faraday synthesis
 \citep{Wolleben2009,Wolleben2010}. The project is subdivided into
 Low- (300--700\,MHz), Mid- (800--1300\,MHz) and High-band (1.3--1.8\,GHz)
 surveys. Observations for the High-band (HB) survey are complete: in the
 north this used the DRAO 26-m, while the southern component (also
 known as STAPS) used the Parkes 64-m telescope.  Early results from the northern survey
 have been published \citep{Wolleben2010b,Sun2015}. Unlike the earlier
 DRAO survey \citep{Wolleben2006}, GMIMS HB fully samples the sky, and
 its multichannel backend gives a good estimate of RM wherever the
 signal is not wiped out by strong depolarization in this band.
 Combined with C-BASS measurements at 5\,GHz this will allow accurate
 extrapolation of the polarization angles to short wavelengths where
 depolarization is negligible.  The second new initiative is the
 S-band Parkes All Sky Survey (S-PASS) at 2.3\,GHz \citep{Carretti2010}. Like GMIMS this is
 a multichannel survey allowing in-band RM measurements, albeit of
 limited accuracy since the available bandwidth is only
 184\,MHz. Observations are complete (STAPS and S-PASS were
 observed commensurately) and initial results were published by
 \citet{Carretti2013}. Although only covering the southern hemisphere,
 S-PASS includes most of the sky regions that are strongly depolarized
 in GMIMS.  As expected, at 2.3\,GHz there is much less
 depolarization, so $RM$s derived from S-PASS and C-BASS should fill
 most of these gaps. In the small fraction of the sky still
 depolarized at 2.3\,GHz in-band measurements using the multi-channel
 southern C-BASS receiver will be used to make the correction.

\begin{figure*}
\centering
\includegraphics[width=0.49\textwidth]{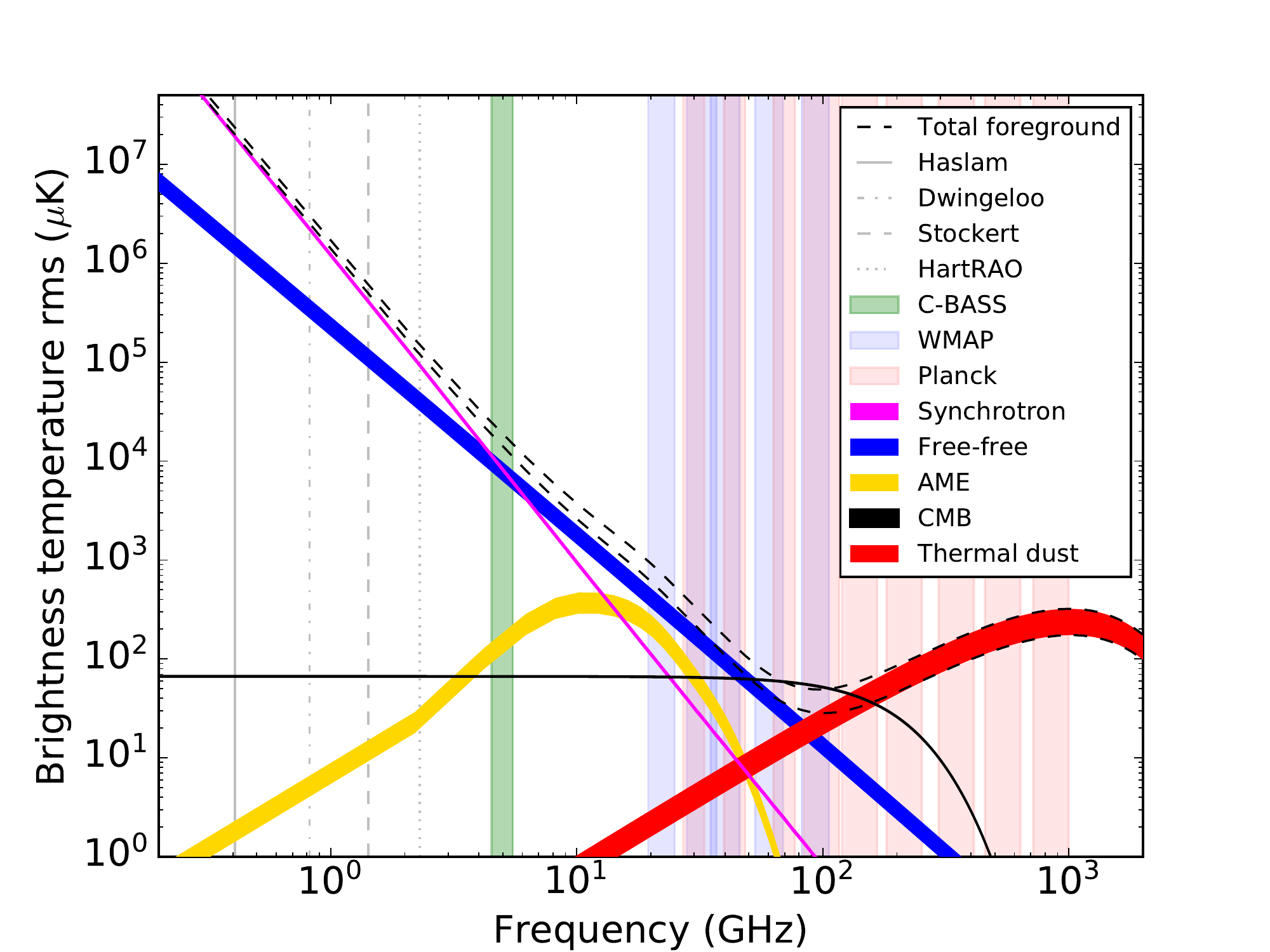}
\includegraphics[width=0.49\textwidth]{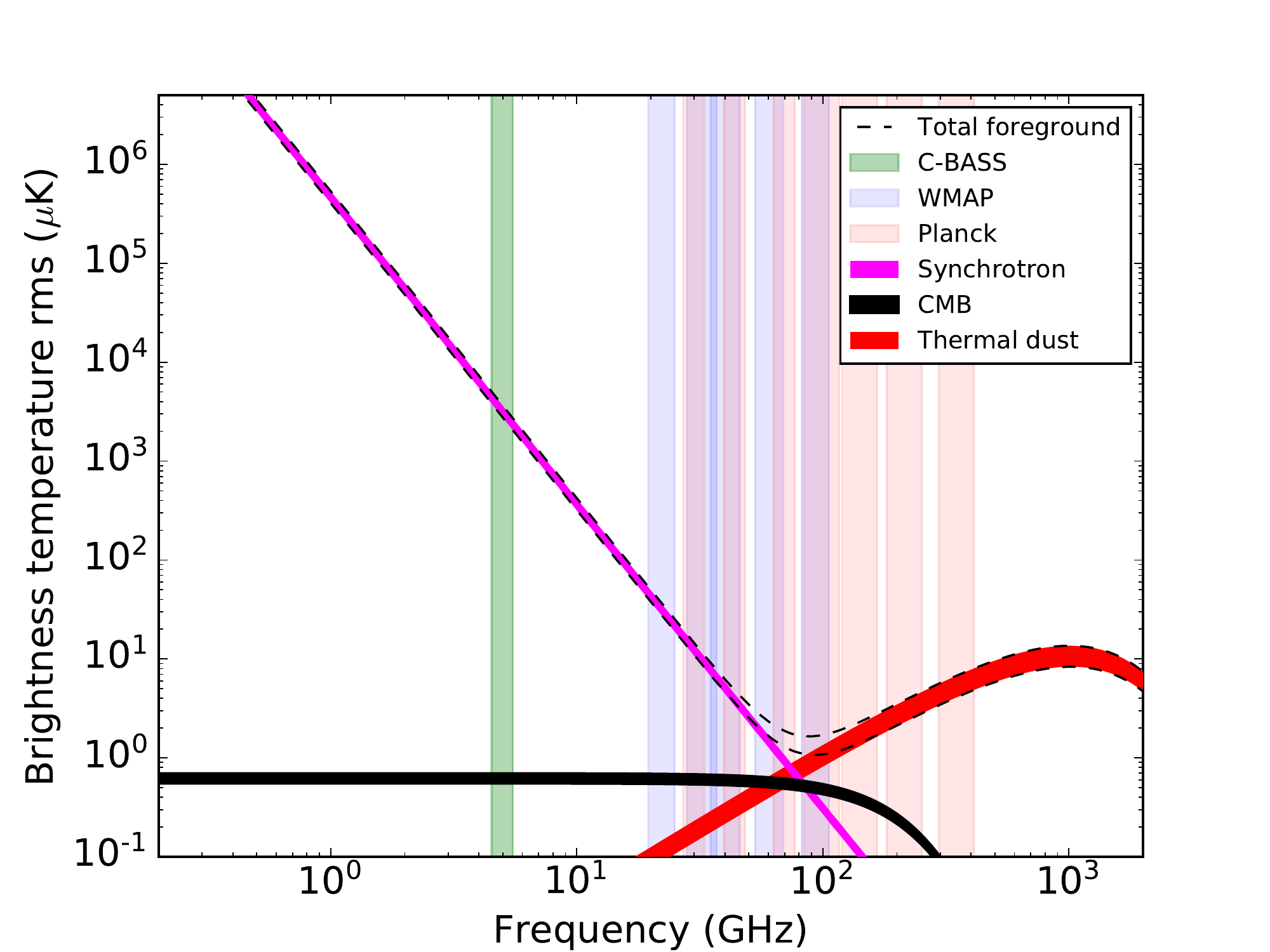}
\caption{Frequency spectra of diffuse foregrounds in temperature ({\it left}) and polarization ({\it right}). Black solid line: CMB temperature and $E$-mode polarization; Magenta line: synchrotron; Blue line: free-free; Red line: thermal dust; Yellow line: anomalous microwave emission; black dashed line: sum of foreground components. The lines indicate the rms fluctuation level in each continuum component from the \citet{Planck2015_X} model (except for the $E$-mode polarization), evaluated at 1\degr\ FWHM resolution, for the region outside the {\it Planck} 2015 HFI Galactic plane masks that include 80 and 90 per cent of the sky (shown by the bottom and top edges of the lines respectively). Underlaid are the bands of \textit{Planck}, \textit{WMAP}, C-BASS, and the lower frequency radio surveys. The $E$-mode polarization amplitude has been taken from \citet{Planck2015_X}, and is calculated from the best-fit power spectrum.}
\label{fig:frequency_spectra}
\end{figure*}

It remains to be seen whether GMIMS and S-PASS will be sufficiently
free of scanning artefacts and far sidelobes to be useful in
constraining the total intensity foreground spectrum. However, such
artefacts are less important for determining rotation measures for two
reasons. Firstly, in the Faraday-thin regime the position
angle-wavelength relation closely follows the simple law:
$\chi(\lambda) = \chi_0 + {\rm RM}\lambda^2$. This allows an internal
consistency check and rejection of outlier data. Secondly, Faraday
rotation causes order unity changes (including sign changes) to the
measured Stokes $Q$ and $U$ parameters. Consequently low-level
artefacts have much less impact than on modeling the Stokes $I$
spectrum, where we are interested in spectral index variations that
may change the intensity ratio between 1.4 and 5\,GHz by 10 per cent
or less.

Between the C-BASS and \emph{WMAP} frequencies, the only large-scale
survey is the QUIJOTE experiment at 11--19\,GHz
\citep{Genova-Santos2015a,Genova-Santos2015b}, which only
covers the northern sky.\footnote{ There are plans, not yet funded, to
  extend the QUIJOTE survey to the southern hemisphere
  (J. A. Rubi\~{n}o-Martin, priv. comm.).}  Unlike C-BASS, QUIJOTE
does not aim to accurately recover very large-scale sky structures,
and it is much less sensitive to the foreground emission, which fades
rapidly with frequency.  However, QUIJOTE does cover the frequencies
over which the anomalous microwave emission rises rapidly to
prominence, and will provide very useful constraints on this
component, especially along the Galactic plane, where component
separation is most complicated. The GEM 5~GHz survey
\cite{Barbosa2006} is at the same frequency and resolution as
C-BASS. It will cover a limited range of declinations in the southern
hemisphere in polarization only (not intensity), and may provide a
useful cross-check on the C-BASS South observations.

%% file: foregrounds.tex
\section{CMB Foregrounds}
\label{sec:mechanisms}

In this section we summarise the properties of the main foreground
components that are known, and review how the new C-BASS data will help with the
problem of cleaning foregrounds from CMB observations. We focus on
`low' frequencies ($\lesssim 100$\,GHz) where synchrotron, free-free,
AME and CMB emissions dominate. At high frequencies ($\gtrsim
100$\,GHz), thermal dust dominates the sky and has been mapped in
detail by new observations from \emph{Planck}
\citep{Planck_Int_XIX,Planck_Int_XXII}, which complement the data from
low-frequency surveys such as C-BASS.

Fig.~\ref{fig:frequency_spectra} shows the frequency spectra of
diffuse foregrounds in intensity and polarization, based on the
modelling by \citet{Planck2015_X}.  At very low frequencies ($<1$\,GHz), synchrotron radiation
invariably dominates due to its steep spectrum, while at higher
frequencies ($\approx 10$--100\,GHz), free-free and AME are
stronger. In polarization, synchrotron dominates up to frequencies of
$\approx 80$\,GHz or higher
\citep{Dunkley2009a,Planck2015_X,Krachmalnicoff2015}. These typical spectra show
that these diffuse components of radiation emit over a similar range
of frequencies with spectra that are hard to discern from each
other. In particular, at frequencies around the peak of the CMB spectrum (150 -- 250 GHz) the spectrum of the CMB is very similar to that of synchrotron emission. Strong spectral lines (e.g., CO and
HCN rotational transitions) can also have a significant impact on the
broad-band intensities measured by the CMB spacecraft
\citep[e.g.,][]{Planck2013_XIII}. The broad-band detectors used in most
CMB experiments cannot distinguish between line emission and the
surrounding continuum, so both components have to be modelled to give
the expected signal in a given frequency channel.

While the total foreground signal is tightly constrained by the
observations, the decomposition into components is currently quite
uncertain, with different model assumptions capable of changing the
ratio of synchrotron to AME power at 30\,GHz by a factor of two
\citep{Planck2015_XXV}. Of course this is one of the main motives for
surveys such as C-BASS, which as we demonstrate in
Section~\ref{sec:impact} will substantially improve the situation.

\subsection{Synchrotron Emission}
\label{sec:synchrotron}

Synchrotron radiation is the dominant low-frequency foreground and
will be the one most constrained by C-BASS.  It is produced by cosmic
ray leptons (electrons and positrons) spiralling in the Galactic
magnetic field \citep{Rybicki_book}. The radio spectrum of a single component of synchrotron radiation is
well approximated by a power-law over a wide range of frequencies, with
brightness temperature $T_B(\nu) \propto \nu^{\beta_{S}}$, which
derives from a power-law distribution of cosmic-ray energies, $N(E)
\propto E^{2\beta_{S}+3}$. Since the local cosmic ray lepton energy
 spectrum is extremely smooth in log-frequency space \citep{AMS2_2014}, and the frequency range of interest
 1.5--150\,GHz maps to only one decade of
 particle energy, the basic synchrotron spectrum is also extremely smooth. However, both intrinsic and line-of-sight
effects can cause the spectrum to deviate from a simple power law,
complicating the process of fitting and removing synchrotron emission
from CMB maps \citep[e.g.][]{2017MNRAS.472.1195C}. Both the observed radio spectrum
\citep[e.g.,][]{deOliveiraCosta2008,Kogut2012}, and direct measurement
of the local cosmic-ray lepton spectrum
\citep[e.g.,][]{PAMELA2011,AMS2_2014} show significant spectral
curvature at a few GHz, corresponding to particle energies of $\sim
5$\,GeV,\footnote{ These energies are near those strongly affected by
  solar modulation of the cosmic ray spectrum, but detailed modelling
  by e.g., \citet{Strong2011} and \citet{DiBernardo2013}, shows that
  the observed curvature is not solely due to solar modulation.}
giving a net change in the spectral index $\beta_S$ from about $-$2.6
at a few hundred MHz to about $-$3.1 above 10\,GHz \citep[e.g.,][]{Strong2011}.  Although spectral
curvature in synchrotron radiation is often attributed to radiative
energy losses, such losses in the interstellar medium cannot explain a
spectral break at this energy, and hence it must be attributed to a
feature in the ill-understood injection mechanism that supplies the
Galactic cosmic ray population. In addition to these causes of
intrinsic spectral curvature, it is expected that on long lines of sight
through the Galaxy, i.e., at low Galactic latitudes, the superposition
of regions with different spectral indices will tend to flatten the
observed synchrotron spectrum at higher frequencies. Observations at
very low frequencies will thus tend to underestimate the synchrotron
contribution at frequencies near to the foreground minimum unless this curvature is taken into account. We can thus expect that multiple
 measurements of the synchrotron component across the microwave band
 will be required in order to determine the spectral shape to the accuracy
 required for future $B$-mode observations.

Our knowledge of the spectrum of intensity of Galactic
synchrotron radiation comes primarily from sky surveys at 0.4\,GHz
\citep{Haslam1982}, 1.4\,GHz \citep{Reich1986}, 2.3\,GHz
\citep{Jonas1998}, and 23 GHz \citep[WMAP;][]{Bennett2003b,Gold2011};
see Table~\ref{tab:surveys}.  Maps of the spectral index across the
sky based on radio total intensity
\citep{Lawson1987,Reich1988,Davies1996,Platania1998,Platania2003,Bennett2003b,Dickinson2009,Gold2011}
and microwave polarization from {\it WMAP}
\mbox{\citep{Fuskeland2014,Vidal2015}} and S-PASS \citep{SPASS2018} show variations in the range
$-4.4 < \beta < -2$.  The flattest spectra are found along the
Galactic plane, and are probably due to free-free emission (and
absorption, at the lowest frequencies). Apparent large-amplitude
variations in spectral index are also found in the regions of weakest
synchrotron emission at high latitudes, which are most susceptible to
the artefacts discussed in Section~\ref{sec:surveys}. The most
reliable maps tend to show the smallest-amplitude
variations. Nevertheless, after correction for the free-free
contribution there is good evidence for genuine spatial variations of
intensity spectral index, with slightly flatter spectra along the Galactic plane
\citep{Planck_Int_XXIII} and in the `haze' near the Galactic centre
\citep{Dobler2008,Planck2013_IX}. Individual supernova remnants (SNR)
and pulsar wind nebulae (PWN), usually taken as the major sources of
Galactic cosmic rays, typically have flatter spectra than the diffuse
synchrotron, from $\beta_S= -2$ to $-2.3$ for PWN
and $-$2.4 to $-$2.8 for shell SNR
\citep{Green2014,Planck_Int_XXXI}. Polarized spectral indices will not necessarily be the same as in intensity, due the summing over different polarization angles within the volume probed by the beam. \citet{SPASS2018} observe that the average spectral index between 2.3 GHz and the 23 -- 33~GHz {\it WMAP} and {\it Planck} bands is $-3.22$ independent of angular scale, but with significant spatial variations that are not simply due Galactic latitude. These variations will complicate efforts to extrapolate synchrotron contamination to the CMB foreground minimum frequencies.

Because synchrotron emission does not dominate the total intensity
foreground in the space microwave band ($\sim 20$--$300$\,GHz), attempts at component
separation have effectively extrapolated it from the most reliable of
the low-frequency templates, i.e. the 408\,MHz survey. This long frequency
baseline, and the poorly quantified variable slope and curvature of
the spectrum, make this one of the main sources of uncertainty in
component separation. The synchrotron-dominated data from C-BASS, at
much higher frequency, will substantially reduce this uncertainty
(e.g., \citealp{Errard2016}).  Further reliable surveys between 5 and 30
GHz would improve the situation even more, as this would tightly constrain measurements of both
the spectral index and spectral curvature as a function of sky position.

\subsubsection{Loops, spurs and the haze}

C-BASS will provide a new look at diffuse Galactic synchrotron
and free-free emission. Given its modest resolution and high brightness sensitivity, this will be 
especially valuable for faint, large-scale structures at intermediate
and high Galactic latitude. Of course, the synchrotron total intensity on these
scales is mapped with high signal-to-noise ratio at 408\,MHz by
\citet{Haslam1982}; however, it is clear from {\it WMAP} and {\it Planck}
that more structure is apparent in polarization; in particular, the
synchrotron loops and spurs are seen with much higher contrast in the 
polarization images \citep{Planck2015_XXV}. 
These features are relatively local, but there may also be a contribution 
from the Galactic halo. Even the weighted average 
{\it WMAP} and {\it Planck} data are not sensitive enough to detect the polarized
emission in the faintest regions, but C-BASS will detect it everywhere, 
and hence address the issue of whether the inter-loop high-latitude emission
is a distinct (e.g., halo) component, in which case it may have a discernibly 
different spectrum, or whether it is produced by numerous overlapping 
structures similar to the visible loops, but fainter.

Of particular interest is
the {\it WMAP}/{\it Planck} haze \citep{Planck2013_IX}, identified as excess
emission at $\approx 1$\,cm partly coincident with the Fermi $\gamma$-ray 
bubbles \citep{Dobler2010,Ackermann2014} which appear to delineate a 10-kpc
scale bipolar outflow from the Galactic centre. The haze is (presumably) 
synchrotron emission with a flatter spectral index ($\beta \approx -2.5$) than the rest of the sky ($\beta \approx -3.0$). However, because of its low
signal-to-noise ratio in the satellite data, and the uncertainty in foreground
separation, it is not clear if the haze is really a distinct component rather than simply a trend to flatter spectral index in the inner Galactic halo, let
alone whether it is related to the bubbles \citep[see e.g.,][]{Planck2015_XXV}.
Including C-BASS in the component separation analysis should pin down
the spectrum of the haze and reveal whether it has a well-defined boundary 
and to what extent it matches the $\gamma$-ray structures.

\subsubsection{Polarized synchrotron and Faraday rotation}
\label{sec:impact-FR}

Optically thin synchrotron radiation has an intrinsic polarization of 70--75 per
cent, oriented perpendicular to the projected magnetic field in the
source region \citep{Rybicki_book}. Although reduced
in practice by superposition of different field directions along the
line of sight, observed polarization fractions can exceed 30 per cent
\citep[e.g.,][]{Vidal2015}. Because these regions may have different
spectral indices, polarized and unpolarized spectra may differ, and
need to be fitted separately. In principle it should be easier to fit
the polarized spectrum, since synchrotron radiation is the dominant polarization
foreground below the foreground minimum; but at present this is
limited by the low signal-to-noise ratio of the {\it WMAP} and {\it
  Planck} polarization maps, and also by large-scale systematic
differences between the two surveys \citep{Planck2015_X} which
indicate residual systematic errors in at least one of them. The C-BASS data will provide the
first measurements of the polarized synchrotron emission that are both high
signal-to-noise ratio, and not affected by depolarization, across most of
the sky.

The Galactic magnetic field reveals itself through both Faraday rotation
and through the intrinsic polarization of the Galactic synchrotron emission, 
which is orthogonal to the projected field direction in the plane of the sky.
Only a band of a few degrees along the plane in the inner quadrants will suffer large depolarization; C-BASS will give a reliable map of projected magnetic field direction at moderate and high latitudes. These lines of sight probe
the local interstellar medium in the plane and the Galactic halo above the 
spiral arms, and so can provide constraints on the measured tangling of the field on relatively
small scales: 1\degr\ corresponds to about 3\,pc for typical structures in
the Population I disc, and $\sim 20$\,pc for a 1-kpc scale-height halo. 
If the halo field is relaxed, the degree of polarization should reach
a substantial fraction of the $m_{\rm max} \approx 75$ per cent expected from 
a uniform
$B$-field; if the structure is tangled, the structure function of the polarized
pattern will give the angular scale(s) of tangling, while random-walk 
depolarization will allow us to estimate the number of reversals on the line
of sight $m \sim m_{\rm max} / \sqrt{N}$; these two approaches give independent
estimates of the tangling scale as a fraction of the scale height. It
will be illuminating to compare the field revealed by synchrotron polarization
with the projected field traced
by dust polarization in emission \citep{Planck_Int_XIX} and absorption \citep[e.g.,][]{Heiles2000,Panopoulou2015}, which give us different weighting functions on the line of sight,
and, for starlight polarization, an upper limit to the distance.
 
At low latitudes the projected magnetic field
is an average along the line-of-sight, but it still gives
information about the field direction and coherence; in fact, modelling
of the magnetic field pattern in the disk hinges on accurate assessment
of the synchrotron fractional polarization at low latitudes, and is currently 
limited by our inability to distinguish synchrotron from AME in the Galactic
disk \citep{Planck_Int_XLII}.

C-BASS data will be combined with 
polarimetry from the GMIMS HB and S-PASS surveys to yield improved maps of 
the Faraday rotation of the diffuse Galactic synchrotron polarization, hence
probing the Galactic magnetic field. Adding
C-BASS doubles the range in $\lambda^2$ compared to GMIMS alone,
yielding a corresponding increase in RM precision, while the precision of 
the intrinsic position angle will be improved by a factor of eight.
Discrepancies between RM values derived in-band from GMIMS
and in combination with C-BASS will reveal breakdown of the simple 
$\lambda^2$ law of Faraday rotation, as expected when there is measurable
variation of Faraday depth across the beam
and/or along the line of sight. Such Faraday dispersion will also 
be associated with depolarization, and so is expected to be seen only around
the borders of regions which are strongly depolarized at the lower
frequency, specifically
at $|b| \la 30\degr$ in the inner quadrants for GMIMS and over a substantially
smaller region for S-PASS  
\citep{Wolleben2006,Carretti2013}. This requires differential rotation of 
$\Delta RM \ga \pi/2\lambda^2$, i.e. $\ga 36$ and 
$92 \rmunit$ at 1.4 and 2.3\,GHz respectively.
Where GMIMS is depolarised (almost exclusively in the southern hemisphere), 
we can derive RM from the combination of S-PASS and 
C-BASS, which increases $\Delta\lambda^2$ by a factor of 5.5 compared to 
using the intra-band $\Delta\lambda^2$ from S-PASS alone.

Similar depolarization at 5\,GHz requires 
$\Delta RM \ga 440 \rmunit$, and hence such depolarization
should be restricted to very low latitudes in the inner Galactic plane 
($|\ell| < 50$). This entire
region will be observed by C-BASS South, and its 128-channel backend 
(8\,MHz channels) will allow us to measure RMs up to $10^5 \rmunit$,
an order of magnitude larger than even that at the Galactic centre 
\citep[$6500 \rmunit$, see][]{Vidal2015}.
(In this region the synchrotron intensity is high enough that it will be
detectable in each channel, except where strongly depolarized.)

The RM map gives a clear look at the line-of-sight structure of the field in the Faraday layer. For example, we would like to know whether it varies smoothly or characterized by abrupt current sheet transitions \citep{Uyaniker2002}. 
When tangential to the line of sight, current sheets show up as 
discontinuities in RM, accompanied by ``depolarization canals''.
It will be particularly interesting to compare the Faraday rotation of the
diffuse synchrotron emission with that of extragalactic sources and discrete
Galactic supernova remnants and pulsars \citep[e.g., ][]{vanEck2011}, 
which will allow us to constrain
models for both the magnetic field geometry and the distribution of emitting
regions along the line of sight \citep{Jaffe2011}.

\subsection{Free-Free Emission} 
\label{sec:free-free}

Free-free emission due to coulombic interactions of electrons with
ions is produced in individual H{\sc ii} regions and the diffuse warm
ionized medium ($T \approx 10,000$~K). The free-free spectrum from a
plasma in local thermodynamic equilibrium (LTE) is accurately known
\citep{Rybicki_book,Draine_book}; in the optically thin regime it has
a near-universal form with spectral index $\beta = -2.1$ at GHz
frequencies, slightly steepening ($\Delta \beta < 0.05$) at
frequencies of tens of GHz and higher.  The steepening slightly
increases as plasma temperature falls, but for the relevant
temperature range the impact is barely detectable.  In contrast, the
transition to the optically thick regime cannot be accurately modelled
at the degree-scale resolution of interest here because it depends on
the brightness distribution within the beam; fortunately this only
becomes a significant issue below $\sim 1$\,GHz, with the brightest H{\sc ii}
regions on the Galactic plane showing absorption effects at 408\,MHz and lower. 

The well-defined spectrum makes free-free emission one of the most
stable solutions in component separation analyses, at least for the
distinct nebulae dominated by free-free emission up to 100\,GHz and
even higher \citep{Planck2014_int_XIV}. In these large H{\sc ii}
complexes, C-BASS data will be dominated by free-free emission, which
will allow verification of the spectral index and provide constraints on
free-free polarization.  On the other hand the diffuse high-latitude
free-free emission is weaker than other foreground components at all
frequencies, making it difficult to separate based on spectral
information alone. Although attempts have been made to use H$\alpha$
templates to constrain models of the high-latitude component
\citep{Dickinson2003,Finkbeiner2003,Draine_book}, for various reasons
this has not proved very accurate \citep{Planck2015_XXV}. Radio Recombination Line (RRL) surveys \citep[e.g.,][]{Alves2015} may also provide an independent and direct tracer of free-free emission.

Free-free emission is inherently unpolarized, but low levels of
polarization (a few percent) can be induced by Thomson scattering
around the peripheries of H{\sc ii} regions \citep{Rybicki_book}, and locally could be
stronger than the synchrotron emission near the foreground minimum
($\nu \approx 70$\,GHz) because of the flatter free-free spectrum; as
yet, this has not been detected.

As we will see in Section~\ref{sec:impact}, C-BASS will dramatically improve
our ability to recover the free-free emission from the Galactic 
warm ionized medium (WIM), including the faint WIM emission at high
Galactic latitudes that is also traced by H$\alpha$. Standard models of
the WIM seem to over-predict the radio free-free emission given the
observed H$\alpha$ \citep[e.g.,][]{Dickinson2003,Planck2015_XXV}, and a more accurate free-free map allowing detailed point-for-point comparison with reasonable
signal-to-noise ratio should help identify the source of the discrepancy, be it unexpectedly
low $T_e$, scattering of H$\alpha$ by high-latitude dust, or departures from
LTE. Because free-free emission comes primarily from H{\sc II} regions, which are strongly clustered with the increased star formation in the Galactic plane, free-free emission dominates the narrow Galactic
plane in the space microwave, and is about equal to synchrotron at 5\,GHz,
as early C-BASS results have shown \citep{Irfan2015}. Here C-BASS will
help recover the spectrum of the subdominant synchrotron emission, which
comes from distant regions of the Galactic disk.

\subsection{Anomalous Microwave Emission (AME)} 
\label{sec:ame}
Anomalous microwave emission is a component of Galactic emission that is strongly correlated with thermal dust emission but has a frequency spectrum that peaks in the tens of GHz  \citep{1996ApJ...460....1K,Leitch1997}; see e.g.
\citet{deOliveira-Costa2004,Davies2006,Gold2011,Ghosh2012, Planck_Int_XV} and \citet{2018NewAR..80....1D} for a review.

AME is clearly seen at 10 -- 60\,GHz with a rising spectrum at low
frequencies and a steeply falling spectrum at higher frequencies,
radically different from the tail of the thermal dust emission, and is
very closely correlated with dust emission at IR/sub-mm wavelengths
\citep{Planck2015_XXV}. The best example comes from the Perseus
molecular cloud where the spectrum has been accurately determined
\citep{Watson2005,Planck2011_XX,Genova-Santos2015b}.  A major problem
for component separation is that the spectrum is spatially variable,
with individual clouds peaking in the range at least 20 -- 50 GHz
\citep{Planck_Int_XV,Planck2015_XXV}. At low latitude we expect
superposition of clouds with a range of peak frequencies, so that AME
can resemble free-free or synchrotron spectra rather closely: along
with the variable synchrotron spectrum this is the second major cause
of the large uncertainty in current component separation.

Measurements of the polarization of AME are challenging due to the
weak signal and difficulties in component separation. Nevertheless, a
number of measurements indicate that AME is at most weakly polarized, with
upper limits of a few per cent in the space microwave band
\citep{Mason2009,Macellari2011,Dickinson2011,Lopez-Caraballo2011,Rubino-Martin2012a,Hoang2013,Planck2015_XXV} and less than 0.5 per cent at lower frequencies \citep{2017MNRAS.464.4107G}.

The source of AME remains uncertain. The leading candidate is electric
dipole radiation from small spinning dust grains
\citep{Draine1998a,Draine1998b}, but another mechanism still in play
is `magnetic dust', i.e., magnetic dipole emission due to thermal
vibrations of ferromagnetic grains, or inclusions in grains
\citep{Draine1999}. Earlier suggestions of hot ($\sim 10^{6}$\,K)
free-free emission \citep{Leitch1997} and flat-spectrum synchrotron
\citep{Bennett2003b}, now seem unlikely due to the peaked spectrum and
close correlation with FIR templates. Spinning dust possibly explains the
low level of polarization and the narrow range of frequencies at which
it is detected.  However, \citet{Tibbs2013} and \citet{Hensley2015}
cite some properties of AME that do not match expectations for
spinning dust, casting serious doubt on this interpretation.

By design, the C-BASS frequency is too low for significant AME to be detected
over most of the sky, which is a major reason why C-BASS substantially
improves the separation of the non-AME components, as the lower space-microwave frequencies can contain both AME and synchrotron emission. If the
peaked spectrum seen in examples such as the Perseus molecular cloud
is typical, AME should be negligible at 5\,GHz and C-BASS will provide
an AME-free template for synchrotron and free-free emission, which in turn will
allow clear identification of actual AME emission at space
microwave frequencies. With an additional low-frequency measurement that is not contaminated by AME, it is possible to break the degeneracy between synchrotron spectral index and AME amplitude (see Section \ref{sec:impact}). Nevertheless, there may be a few lines of sight where AME is
detectable, allowing \mbox{C-BASS} to constrain models of the low-frequency tail of its
spectrum; a good example is G353.05+16.90 ($\rho$~Oph West) on $1^{\circ}$ scales, where there may still be appreciable AME at 5\,GHz \citep{Planck2011_XX}. If any of the dust-correlated features so
evident in the {\it WMAP} and {\it Planck}-LFI maps are visible in
C-BASS, this could imply a radically different emission mechanism from
spinning dust.

\subsection{Thermal Dust}
\label{sec:dust}

Interstellar dust grains, with sizes ranging from a few to several
hundred nanometers, absorb optical and UV starlight and re-emit via thermal
vibrations in the crystal lattice, which excite electric dipole
radiation \citep{Draine_book}. This is the
dominant foreground above 70\,GHz. Dust emission can be fitted with a modified blackbody, i.e.,  a Planck spectrum, $B(\nu,T_{\rm d})$ multiplied by an emissivity $\propto \nu^{\beta_{\rm d}}$.  The latest {\it Planck} fits to the spectrum below 1\,THz \citep{Planck2015_X} give a narrow range around $\beta_{\rm d} \approx 1.53$, with an rms of 0.03 that may be dominated by fitting errors; $T_{\rm d}$ ranges from 15--27\,K, with a mean $\approx 21$\,K and a standard deviation of 2.2\,K.  However this model over-predicts the data above 1\,THz, where the best-fit values are $\beta_{\rm d} \approx 1.50$ and 
$\langle T_{\rm d} \rangle \approx 19.6$\,K
\citep{Planck_Int_XXII}. 

The apparent uniformity of the dust spectrum disguises considerable spatial
variation in dust properties. \citet{Planck_Int_XVII} showed that $T_d$ 
is anticorrelated with emissivity at high Galactic latitude, the opposite of
what would be expected from variations in starlight intensity, 
implying significant
variations in the UV/optical absorption to FIR emission ratio. There are at
least two, and likely more, chemically-distinct grain populations 
\citep{Draine_book}.  There are certainly
real spatial variations in $\beta_D$; for instance, the Small
Magellanic Cloud has $\beta_D \approx 1.2$ \citep{Planck2011_XVII}. 
Laboratory-synthezised grain analogues show a range
of $\beta_D$ and also spectral curvature \citep{Coupeaud2011}, and the observed
mm-wave spectrum presumably represents whatever reasonably abundant grain 
population has the slowest fall-off towards long wavelengths. 

Polarization of dust emission is due to anisotropic optical properties
of the grains and a preferred orientation with respect to the magnetic
field. Polarized optical extinction is associated with silicates
\citep{Draine_book}, which
are also believed to dominate the mm-wave dust emission
\citep[e.g.,][]{Planck_Int_XXIX,Fanciullo2015}, and, as expected, 
the polarization angles seen in emission and absorption are strongly correlated
\citep{Planck_Int_XXI}.
The intrinsic polarization fraction of thermal dust emission may be around 26
per cent \citep{Planck_Int_XLIV}; as for synchrotron radiation this is reduced
by geometric depolarization, but observed polarization can reach 20
per cent, with typical values of $\approx 5$ per cent
\citep{Planck_Int_XIX}. Also as for synchrotron radiation, these effects can lead
to different spectra in polarization and total intensity, and in fact
the polarized spectrum is slightly steeper \citep{Planck_Int_XXII}.

The intrinsic complexity of the dust spectrum poses a challenge for observing strategies that concentrate on frequencies above 100 GHz. Although synchrotron emission is below the dust emission in this frequency range, without effective constraints on the synchrotron spectrum, degeneracies between different dust models and residual synchrotron will compromise the accuracy of foreground separation at the levels of precision needed for accurate $B$-mode measurements. Although C-BASS measures frequencies far from the peak of the dust spectrum, removing these degeneracies in component fitting can lead to improvements in the measurements of the dust parameters through the improved fitting of the other components (see Section \ref{sec:impact}).

%% file: requirements.tex

\setlength{\tabcolsep}{3pt}
\begin{table}
\caption{Key specifications of the C-BASS survey.}
\label{tab:overview}
\begin{tabular}{l|cc}
\hline
                     &North     &South   \\
\hline
Location            &OVRO       &Klerefontein  \\
                    &California &South Africa\\
Latitude           &37\degr\ $14'$N, & 30\degr\ $58'$S,\\
Longitude           & 118\degr $17'$W & 21\degr $59'$E  \\
Telescope           &6.1\,m Gregorian &7.6\,m Cassegrain \\
Sky Coverage        & $\delta > -15\fdg 6$ & $\delta < 28\fdg 6$ \\
Frequency range     &\multicolumn{2}{c}{4.5 -- 5.5\,GHz}   \\
Effective centre frequency & 4.783\,GHz & 5\, GHz \\
Effective bandwidth           &0.499\,GHz          &1.0\,GHz   \\
Frequency channels   & 1       & 128  \\
Angular resolution   &\multicolumn{2}{c}{$45$ arcmin FWHM} \\
Stokes coverage      &\multicolumn{2}{c}{$I, Q, U (V)$}                     \\ 
Sensitivity          &\multicolumn{2}{c}{$\lesssim 0.1$\,mK r.m.s. (per beam)} \\
\hline
\end{tabular}
\newline 

\end{table}
\setlength{\tabcolsep}{6pt}

\section{Survey requirements and constraints}
\label{sec:requirements}

The resolution requirement of the C-BASS survey is partly set by that of the
complementary surveys at other frequencies and partly by the
 science goals, but it is also limited by practical
constraints. {\it WMAP} and {\it Planck} have resolutions at their lowest
frequencies of $\approx 48$ arcmin and $\approx 33$ arcmin respectively, while the 408~MHz Haslam
et al. map has a nominal resolution of $51$ arcmin. In order to remove foregrounds
at the angular scale of the peak of the $B$-mode power spectrum at
$\ell \approx 90$, a resolution of around 1\degr\ is required. The
resolution is also ultimately set by the size of antenna available,
and the need to under-illuminate it to minimise sidelobes.  With a
6.1-m antenna available, it was possible to design for a beam FWHM of
$45$ arcmin. This is slightly better than the resolution of the Haslam
map and sufficient to clean CMB maps well in to the region of the
$B$-mode power spectrum peak.

Ideally C-BASS would detect polarized emission across the entire sky. To
estimate the level of polarized emission at high Galactic latitudes,
and hence the sensitivity required, we extrapolated from the {\it WMAP} K-band
polarization map. Assuming a mean temperature spectral index of $\beta
= -3$, we estimate that the polarized intensity at 5\,GHz will be
greater than 0.5\,mK over 90 per cent of the sky. We therefore set a
sensitivity goal of 0.1\,mK per beam in polarization. This corresponds to about 14\,mJy in flux density sensitivity. At this
sensitivity level the C-BASS intensity map will be
confusion limited. We estimate the confusion limit from the source counts in the GB6 survey \citep{GB6}, which can be modelled as $N(S){\rm d}S = 76 \, (S/{\rm Jy})^{-2.44}\,{\rm Jy ^{-1}\, sr^{-1}}$.  With  a beamsize of $45$ arcmin the expected confusion
limit from extragalactic sources is about 85\,mJy, corresponding to 0.6\,mK, for an upper flux density limit of 100\,mJy (roughly the individual source detection level in C-BASS maps). In practice the confusion limit will be somewhat lower than this, since the source counts are known to flatten at lower flux density levels than the lower limit of GB6. The
polarization maps will not be confused, as the typical polarization
fraction of extragalactic sources is only a few per cent. It will also
be possible to correct the C-BASS intensity maps for source confusion
using data from higher resolution surveys such as GB6 and PMN \citep{PMN}. The overall specifications of the C-BASS survey are summarized in Table \ref{tab:overview}.

\subsection{Survey Design}

In order to map the entire sky with sensitivity to all angular scales
up to the dipole, the only feasible instrument architecture is a total
power scanning telescope. An interferometer is not feasible because of
the difficulty in obtaining information on scales larger than the
inverse of the shortest baseline. To cover the entire sky from the
ground required two instruments, one in each hemisphere, situated at
latitudes that give significant overlap in the sky coverage to ensure
continuity on large scales between the two halves of the survey and good 
cross-calibration. We also require sensitivity to both intensity and polarization.

In order to construct a sky map with good accuracy on large angular
scales we require a scan strategy with long continuous sweeps of the
sky and good cross-linking of scans (i.e., each pixel is crossed by
several scans in different directions). For intensity measurements we
also choose to use a fixed reference temperature rather than a
differential measurement that switches out signal at the separation
angle between the beams. We scan at constant
elevation to minimise the variation in atmospheric
emission and ground spillover during a scan, . The survey strategy is therefore to make constant-elevation
scans over the entire azimuth range, at the maximum slew rate that the
telescope can manage. Maximising the slew rate pushes the signal
frequency band in the time-ordered data as far as possible away from
any residual $1/f$ noise in the receiver noise power spectrum. The
fastest convenient azimuth slew rate for both C-BASS telescopes is 4
deg/sec. We actually use several different slew rates close to 4
deg/sec so that any systematics in the data that are at fixed
frequency (for example, related to the receiver cold head cycle
frequency or the mains frequency) do not always map to the same angular
scale on the sky. The telescope is slewed at full speed from 0\degr\ to 
360\degr\ azimuth, and then decelerates, halts, and turns
around. This gives a small region of overlap in azimuth coverage and
ensures the whole sky is covered at full slew speed. We also have full
sky coverage in both clockwise- and anti-clockwise-going scans. 

\begin{figure}
\includegraphics[width=\columnwidth]{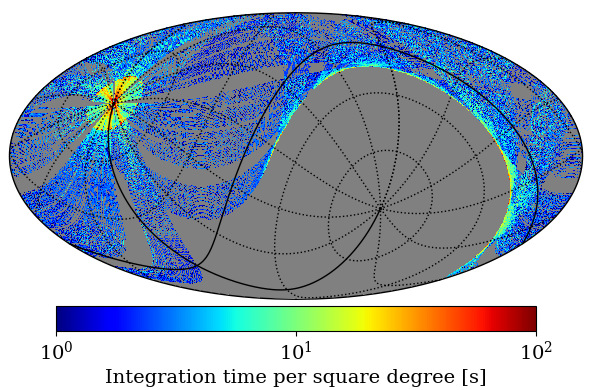}
\includegraphics[width=\columnwidth]{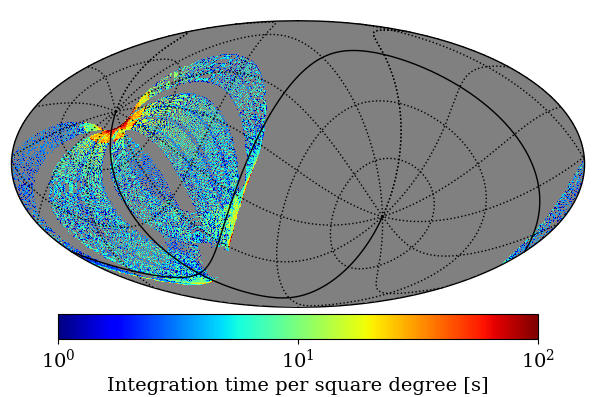}
\includegraphics[width=\columnwidth]{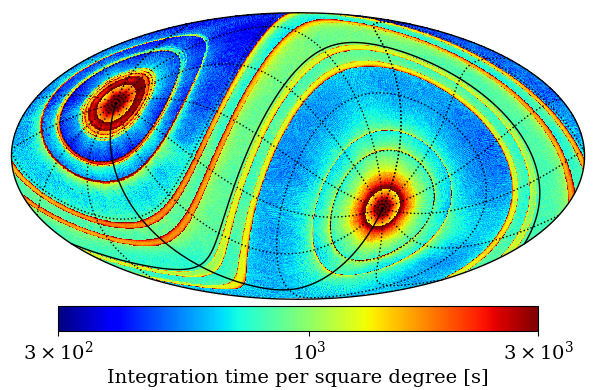}
\caption{{\it Top}: Sky coverage from roughly one day of observations with C-BASS north, using scans at a single elevation going through the north celestial pole (elevation $37\degr$). The map is in Galactic coordinates, with an equatorial co-ordinate grid overlaid. {\it Middle}: Sky coverage from scans at an elevation ten degrees above the celestial pole (elevation $47\degr$), showing how these scans fill in the sky coverage at mid declinations. {\it Bottom}: Complete sky coverage expected from northern and southern surveys combined, using data from all elevations. \label{fig:scans}}
\end{figure}

Scanning at constant elevation equal to the latitude of the observing
site $\phi$ results in the scans always passing through the
celestial poles, and the entire sky is eventually covered down to
declination $\delta = -90\degr+2\phi$ (in the northern
hemisphere). Scanning through the pole has the additional benefit that
the same point on the sky is observed every scan, giving an immediate
check on the drifts in offsets due to the atmosphere of the
receiver. However, the resulting sky coverage is very non-uniform,
with deep coverage at the pole and at the lower declination limit, but
much sparser coverage at intermediate declinations. In order to get
sufficient integration time over the whole sky we also observe at
higher elevations, with about 60 per cent of the survey time spent at
the elevation of the pole and decreasing amounts of time spent at 10,
30 and 40 degrees above the elevation of the pole. This results in a
much more uniform sky coverage (see Figure
\ref{fig:scans}).

For scans at a given elevation, any residual ground spillover signal will
be a fixed function of telescope azimuth. The azimuth at which any
given declination on the sky is observed is also fixed (in fact each
declination is observed at two azimuths, symmetrically placed about the
meridian), which means there is a degeneracy between the
ground spillover and the sky for sky modes that are circularly symmetric
about the pole (these are the $m = 0$ modes in the spherical harmonic
decomposition of the sky in equatorial coordinates). This degeneracy
can be partly broken by observing at different elevations, which have
somewhat different ground-spillover profiles, and by using the overlap
region between the northern and southern surveys, which will have
quite different ground-spill profiles.
 With the northern telescope at latitude $\phi = +37\degr$
and the southern telescope at latitude $\phi = -31\degr$ the overlap
region between the two surveys is from declination $\delta = +28\degr$
to $\delta = -16\degr$. This overlap region also allows for extensive
calibration cross-checks between the two surveys. 

The telescopes observe continuously day and night,
with calibration observations (including sky dips) inserted roughly every
two hours. No attempt is made to synchronize scans, as the sky is covered 
many times in the course of the survey observations.
Contamination from the Sun or Moon is
assessed after the observations, and the final survey data will be
tested empirically for residual contamination. This gives us the
maximum freedom to include good data, but the survey timing is planned
such that even using strictly night-time only data will give
sufficient integration time.

%% file: design.tex
\section{Instrument design}
\label{sec:design}

\subsection{Overview}

\begin{figure}
 \centering
 \includegraphics[width=0.45\textwidth]{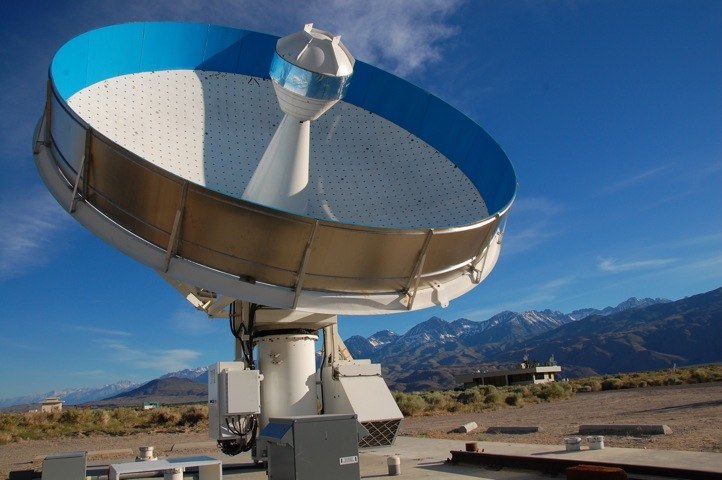}
 \includegraphics[width=0.45\textwidth]{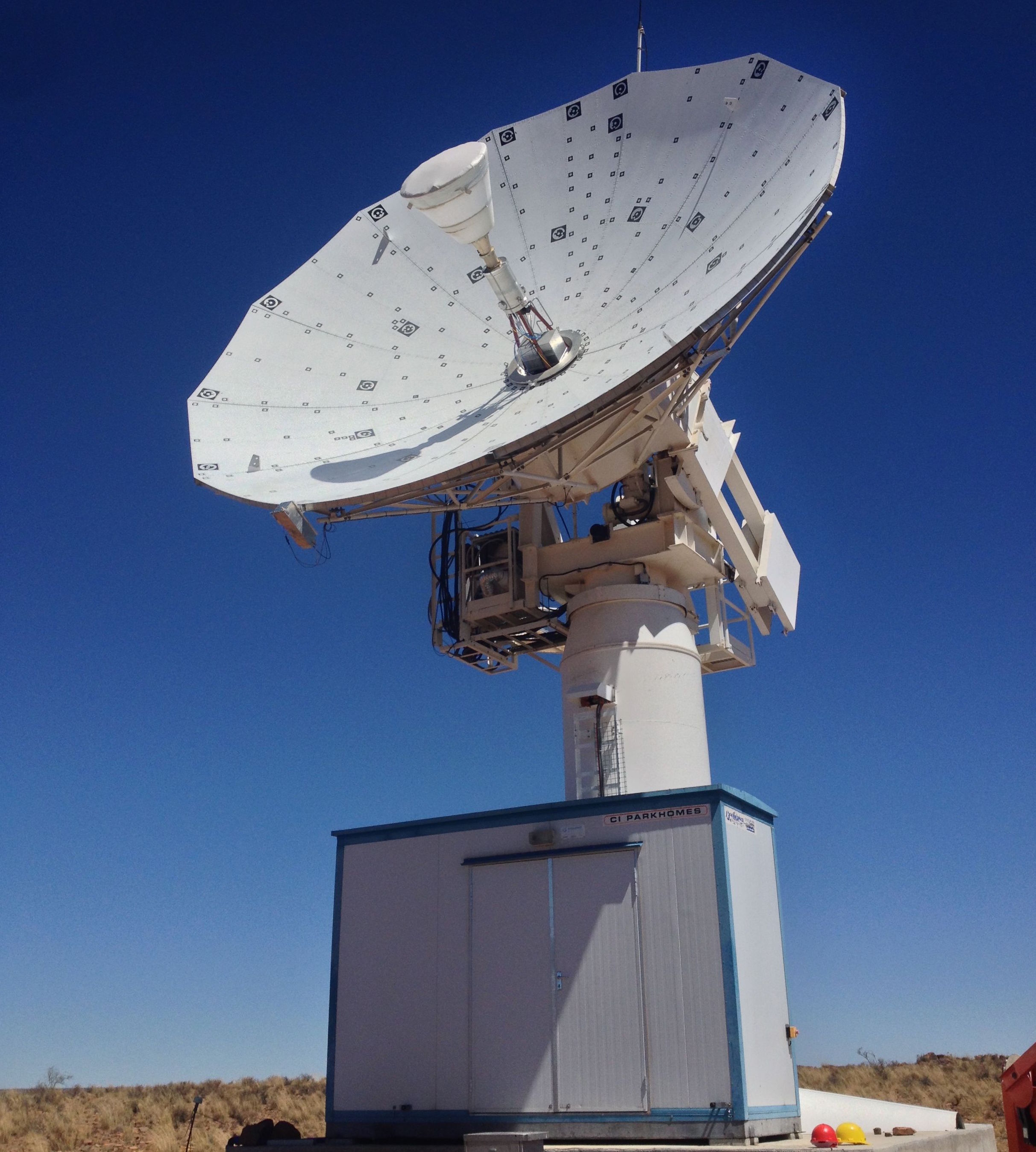}
 \caption{{\it Top}: The C-BASS North telescope, located at the Owens Valley Radio Observatory in California, U.S.A. {\it Bottom}: The C-BASS South telescope, located in the Karoo desert, South Africa. The weather shield around the receiver is removed in this image, showing the lower part of the feed horn and cryostat.}
 \label{fig:telescopes_picture}
\end{figure}

The two C-BASS systems, north and south, have been designed to produce
a single unified survey, and have many features in common. However
there are some significant differences in implementation between the
two systems, some forced by practical constraints, and others due to
improvements in technology and lessons learned between the northern
system, which was designed first, and the southern system. The two
telescopes (see Figure \ref{fig:telescopes_picture}) are similar in
size but differ in numerous details. The northern telescope was
donated to the project by the Jet Propulsion Laboratory, having been
designed as a prototype for an array element for the Deep Space
Network \citep{Imbriale2004}. It has a 6.1-m single-piece reflector with
focal ratio $f/D = 0.36$. The southern telescope was donated by Telkom
SA to SKA South Africa and was originally designed for the ground
segment of a low-earth orbit telecommunications satellite
constellation. It has a segmented 7.6-m primary with twelve radial
panels, and also has a focal ratio of $f/D=0.36$. However, since the
same area of the primary is illuminated as on the northern
antenna, i.e. a 6.1 m diameter, the effective focal ratio of the
southern antenna is 0.46. This difference results in our having to use
different optical configurations for the two telescopes -- the
northern antenna uses Gregorian optics, while the southern antenna
uses Cassegrain optics. Nevertheless, the two antennas have very
well-matched beams \citep{Holler2011}. The northern receiver is an
all-analogue system \citep{king2014}, while the southern receiver
(Copley et al., in prep.) implements the same architecture with a digital
back-end that also provides spectral resolution within the band.

\subsection{Optics}

A total-power scanning telescope is vulnerable to scan-synchronous
systematics, i.e., spurious signals appearing in the time-ordered data at
the same frequency as astronomical signals. The most obvious cause of
such contamination is pick-up of the ground and other non-astronomical
sources of radiation in the sidelobes of the antenna. To mitigate this, we
have designed the optics to minimize the far-out sidelobes as much as possible. This is
achieved by designing an optical system with minimal blockage and
scattering, and very low edge illumination. Full details of the optical design are given by \citet{Holler2011}. Given that we only had
on-axis telescopes available, we were constrained to use a blocked aperture,
rather than an off-axis unblocked design. The secondary mirror blockage results in unavoidable near-in sidelobes, which can however be quite accurately modeled and measured, and hence corrected for in the map analysis. Far-out sidelobes were minimized
by having the secondary mirror supported on a transparent
dielectric material rather than using metal struts. This also has the effect of
maintaining the circular symmetry of the optics and thus minimizing
cross-polarization. We also used a feed horn with very low
sidelobes, which minimizes direct coupling between the feed and the
ground when the telescope is pointed to low elevations.

 The feed is a profiled corrugated horn that generates
 HE12 modes in a cosine-squared section, which are phased up with the
 dominant HE11 mode in a cylindrical final section, resulting in a
 beam pattern with very low sidelobes and cross-polarization. In both
 telescopes the feed is well forward of the dish surface, and the
 entire receiver assembly is mounted above the dish surface. The
 feed to subreflector distance is less than 1\,m in each case, which
 allows the subreflector to be mounted off the receiver assembly using
 a structure made of Zotefoam Plastazote, a nitrogen-blown polyethylene foam. This foam has
 very low dielectric constant and RF losses, and allows the
 subreflector to be supported without the use of struts that would
 cause scattering and break the circular symmetry of the antenna.

To minimize far-out sidelobes and hence reduce ground pick-up, the
northern telescope has absorptive baffles around the primary and
secondary mirrors. The primary baffle intercepts radiation that would
otherwise spill over the side of the dish to the ground, while the
secondary baffle reduces direct radiation from the feed to the
sky/ground. Although these baffles increase the temperature loading on the
receiver and contribute to the system temperature, they significantly
reduce the scan-synchronous ground pick-up. The southern telescope has
a larger (7.6-m) primary and so, when illuminated to produce the same
beam size as the 6.1-m northern telescope, has extremely low edge
illumination and negligible spillover lobes. The Cassegrain design of
the southern telescope means that a baffle around the secondary mirror
is not possible.

Even better rejection of ground pick-up could be achieved by
surrounding the telescopes with a reflecting ground screen that
shields the horizon. This would mean that the environment seen by the
telescope is all at the temperature of the sky, which is around two
orders of magnitude colder than the ground.  Unfortunately the large
size of ground screen required to shield the telescopes and still
allow access to a reasonable range of elevations on the sky was too
expensive to build.

\subsection{Radiometer and polarimeter}

The C-BASS receivers \citep[][Copley et al., in prep.]{king2014}
measure both intensity and linear polarization.
The intensity measurement uses a continuous-comparison radiometer, which compares the power
received by the antenna to a stabilized load signal, using the same gain chain for both signals so
that gain instabilities in the electronics can be effectively
removed. The same basic design has been used in
previous instruments such as the {\it Planck} Low Frequency Instrument
\citep{Bersanelli2010}. In this design, a four-port hybrid is used to form
two linear combinations of the feed and reference signals, which are
then both amplified, before being separated with a second hybrid and
the powers of each signal detected and differenced. Gain fluctuations
in the amplifiers affect both feed and reference signals equally, and
are therefore cancelled out. This cancellation is continuous and does
not rely on a switching frequency, and is more efficient than a Dicke
switch \citep{Dicke1946}, in which half the integration time is spent
looking at the reference load. To protect against
gain fluctuations in the detectors, which come after the sky
and load signals have been separated, phase switches are introduced
in to the two gain arms. A single ideal 180-degree phase switch in one
arm will cause the feed and reference signals to swap between the two
detectors, allowing cancellation of detector gain differences. Non-ideal
performance of the phase switch (e.g., different gains in the two phase
states) are cancelled out by placing phase switches in both
arms, and cycling between all four states of the two switches.

Polarization is measured by taking the complex correlation
of the right and left circular polarizations, which yields $Q$ and $U$
directly as the real and imaginary parts of the correlation:
\begin{eqnarray}
\langle |E_R|^2 + |E_L|^2 \rangle &=& I \\
\langle E_R E_L^* \rangle &=& (Q + iU)/2 \\
\langle E_L E_R^* \rangle &=& (Q - iU)/2 \\
\langle |E_R|^2 - |E_L|^2 \rangle &=& V 
\end{eqnarray}
where complex amplitudes $E_{R,L} = (E_x \pm iE_y)/\sqrt{2}$ 
multiply the propagator $\exp[i(kz - \omega t)]$
\citep{Hamaker1996b}. 
This means that $Q$ and $U$ are measured simultaneously and continuously,
without needing any polarization modulation or physical
rotation. This is more accurate than taking either the difference in power of the individual linear polarizations, or correlating linear polarizations, both of which require subtracting quantities involving the total intensity in order to obtain the much smaller linear polarization signal. Intensity fluctuations in the right and left channels from
the unpolarized atmospheric background, and from the low-noise
amplifiers, are uncorrelated and appear in the $Q$ and $U$
measurements only as noise terms. 
Stokes $V$ can in principle be obtained from the difference of the intensities in right and left circular polarization (Eqn. 4). However, astronomical circular polarization is expected to be extremely small, and accurate measurement of $V$ would require very precise calibration of the individual intensity measurements. In practice the $V$ signal is used as a check of the relative calibration of the intensity channels. 

\subsection{Cryogenic receivers and analogue electronics}

The receivers for the two C-BASS telescopes are similar but differ in
some significant details (\citealp{king2014}, Copley et al., in prep.). The
cryostat bodies are very similar, and both use
two-stage Gifford-McMahon coolers. The northern receiver uses a
Sumitomo Heavy Industries (SHI) SRDK-408D2 cold head, which cools the second stage to 4~K. The southern receiver uses
an Oxford Cryosystems Coolstar 6/30 cold head, which cools to 10~K. The southern cold head does not reach such a cold
base temperature but uses significantly less compressor power (3\,kW vs
9\,kW for the SHI system). 

\begin{figure*}
\includegraphics{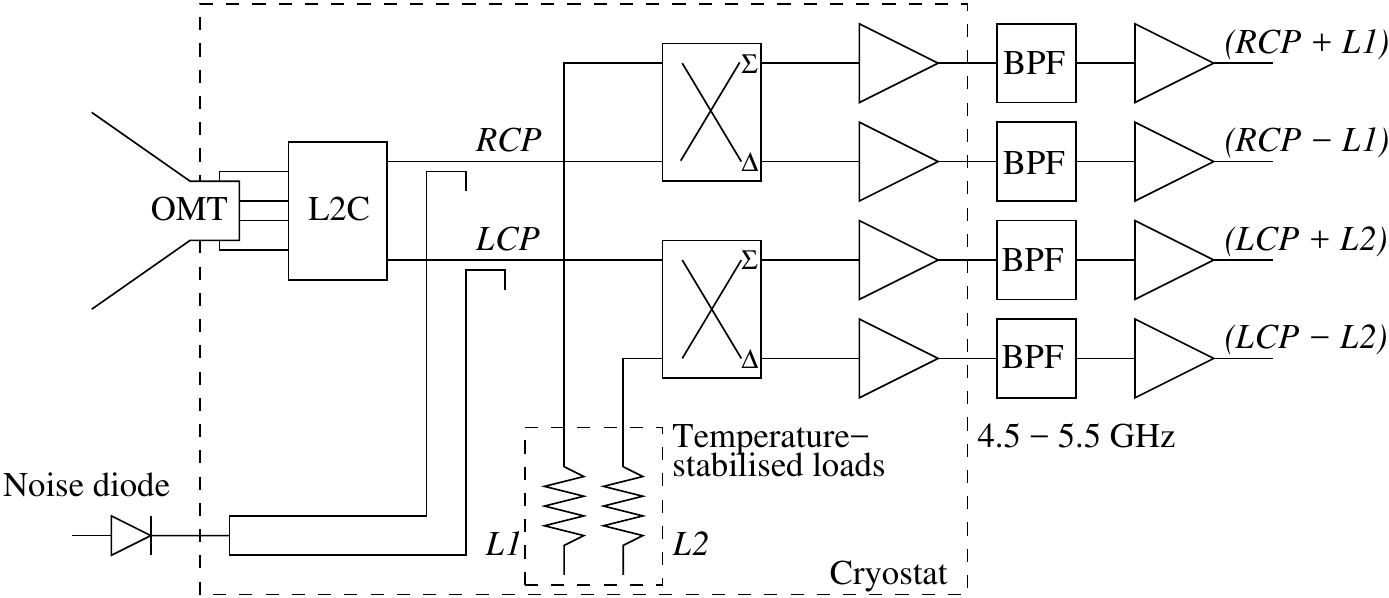}
\caption{Simplified block diagram of the C-BASS front end, which is
  common to C-BASS north and south.  Key: OMT = orthomode transducer,
  L2C = linear to circular converter, $\Sigma, \Delta$ = sum,
  differencing, BPF = bandpass filter, RCP = right circular
  polarization, LCP = left circular polarization. L1 and L2 are
  matched loads. \label{fig:block-rx}}
\end{figure*}

Both receivers use the same design of corrugated feedhorn. The main
body of the feedhorn is at ambient temperature and is bolted directly
to the cryostat body. The upper section of the feedhorn also provides
the support for the secondary mirror assembly. The smooth-walled
throat section of the horn is machined directly into the first-stage
heat shield of the cryostat, and the orthomode transducer (OMT) is mounted
onto the second-stage cold plate. The 4-probe OMT \citep{grimesOMT} is connected via
coaxial cables to a planar circuit that combines the linearly
polarized signals and produces circularly polarized outputs. Coaxial
$-30$~dB directional couplers are used to couple in the noise source
signal used for calibration. The circularly polarized signals are
combined with reference signals in two 180\degr\ hybrids. The reference
signals are generated from temperature-stabilized matched loads
controlled by an external PID controller, which provide a load
temperature stable to better than 1\,mK (see Figure \ref{fig:block-rx}). 

Both receivers use LNF-LNC4$\_$8A low noise amplifiers from Low Noise
Factory, which provide 40\,dB of gain between 4 -- 8\,GHz with a typical
amplifier noise temperature of 2 -- 3\,K. In the southern system the
signals then simply leave the cryostat via stainless steel cables. In
the northern system there are notch filters that remove ground-based
RFI near the centre of the band, reducing the effective bandwidth in polarization from 1 GHz to 499 MHz, and shifting the effective centre frequency to 4.783~GHz.

\subsection{Backends and readout}

\begin{figure*}
\includegraphics[width=0.8\textwidth]{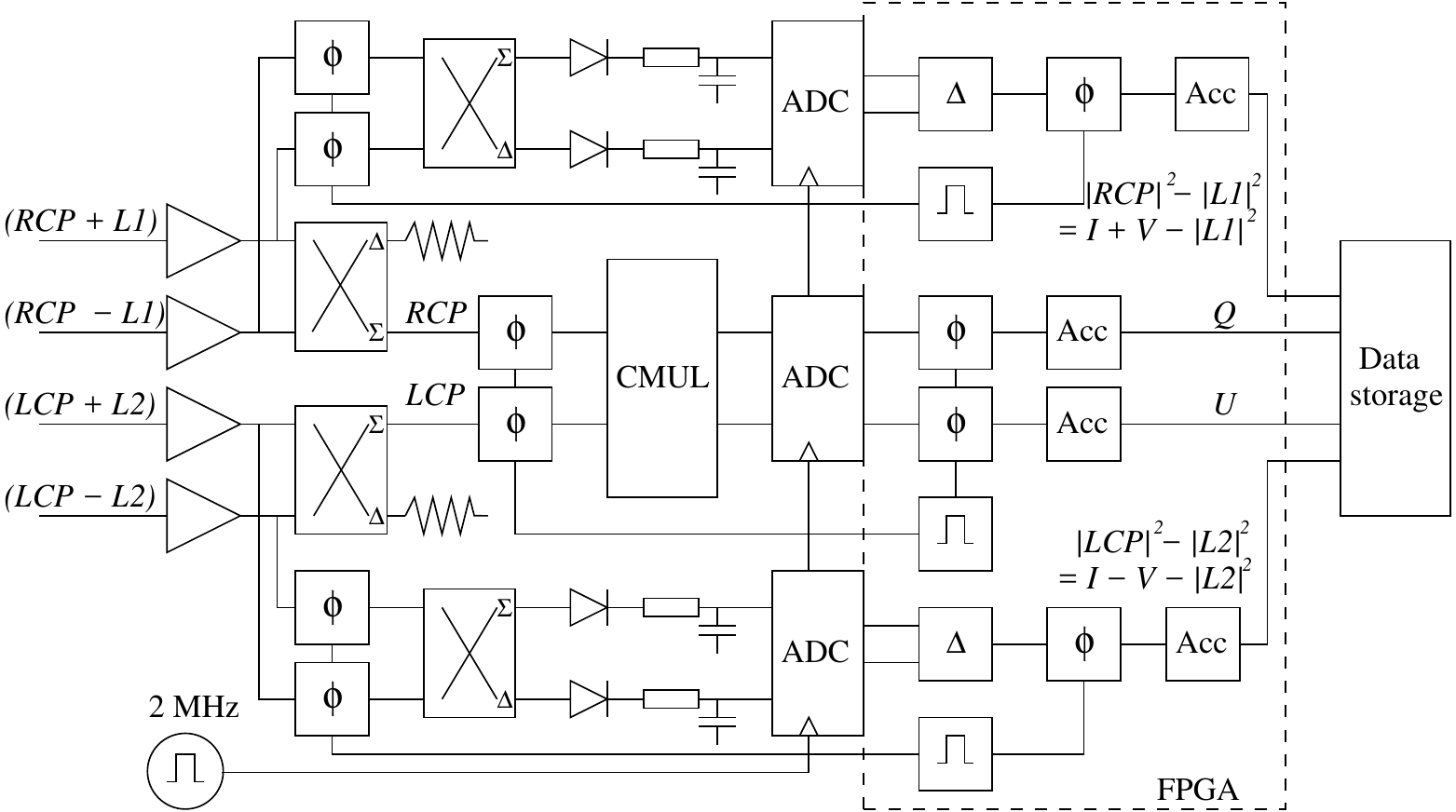}
\vspace{5mm}
\includegraphics[width=\textwidth]{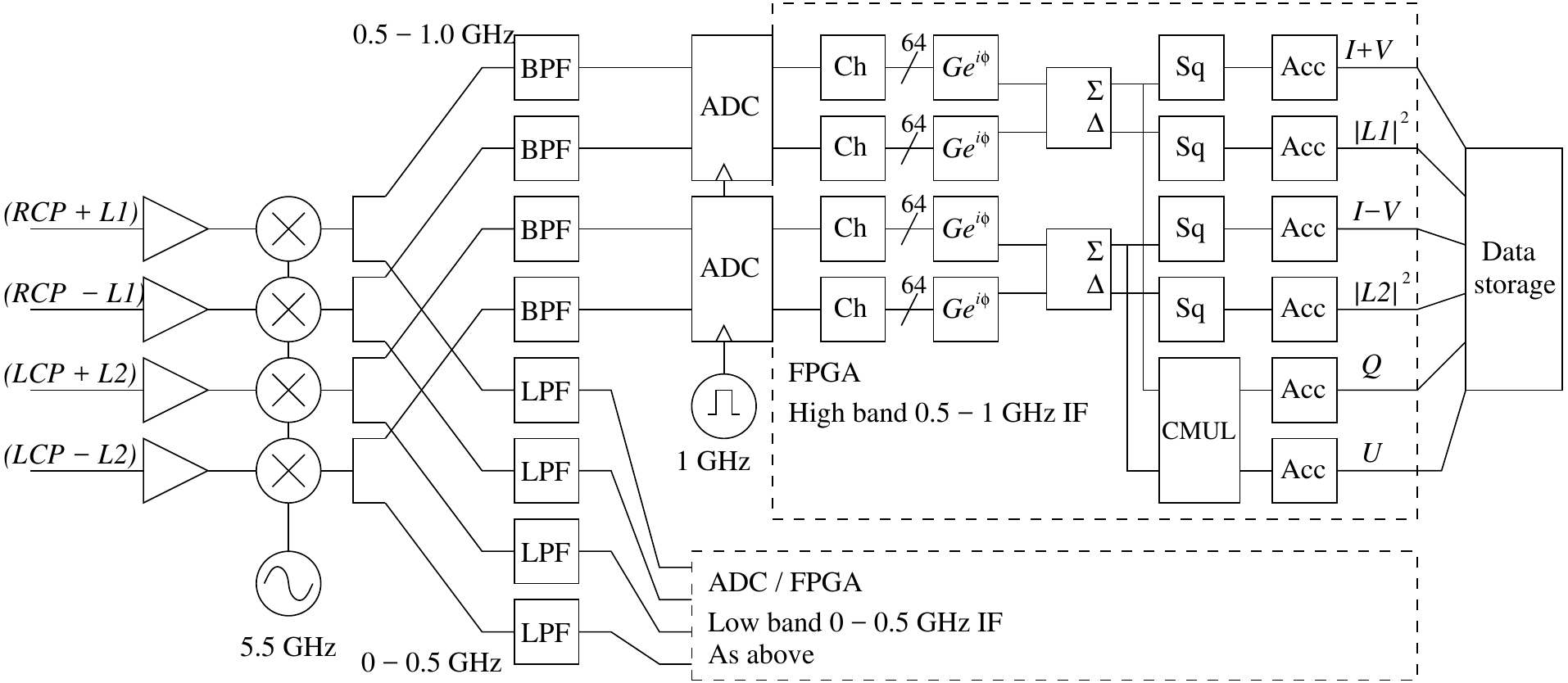}
\caption{Block diagrams of the C-BASS radiometer/polarimeter
  systems. {\it Top}: C-BASS north analogue backend, {\it Bottom}: C-BASS south
  digital backend. Key: $\phi$ = phase switch modulation/demodulation,
  $\Sigma, \Delta$ = sum, differencing, ADC = analogue to digital
  converter, Acc = accumulator, CMUL = complex multiply, BPF =
  band-pass filter, LPF = low-pass filter, Ch = channeliser, $G\rm
  e^{i\phi}$ = complex gain correction, Sq = square detector
  (evaluates $VV^*$ on complex voltages $V$).\label{fig:block-pol}}
\end{figure*}

The two C-BASS receiver systems implement the same signal processing
operations to generate the intensity and polarization measurements,
but in very different ways (see Figure
\ref{fig:block-pol}). The northern system is described in detail in
\cite{king2014}. The radiometer and polarimeter functions are
implemented by analogue electronics operating on the whole RF band as
a single channel. The radiometer uses 180\degr\ hybrids identical to
those used in the cryostat to separate out the sky signal from the
reference signals, which are then detected with Schottky diodes. Phase
switches in the RF signal path cause the sky and reference signals to
be alternated between the physical channels, averaging out any gain
differences or drifts in the amplifier and detector chain. The data
are sampled at 2 MHz following post-detection filtering to 800 kHz
bandwidth, and the sky and reference signals are differenced before
phase switch demodulation and integration to 10 ms samples. For the
polarimeter operation, the separated sky signals are correlated using
a complex analogue correlator consisting of 90\degr\ hybrids and
detector diodes. Again phase switching is used to ensure gain
differences do not bias the correlated outputs.  The detector diode
outputs are filtered, sampled, synchronously detected at the phase
switch frequencies, and filtered and averaged down to 10~ms samples in
an FPGA.

The southern system, by contrast, is fully digital. After further gain
and bandpass filtering, the four RF signals from the cryostat are
downconverted using a 5.5\,GHz local oscillator to an IF band of 0 --
1\,GHz. The lower sideband is used to ensure that images of strong
out-of-band signals from geostationary satellites in the range 3.5 --
4.5 GHz are not aliased in to the IF bands. The IF signals are then
split and filtered to give 0 -- 0.5 and 0.5 -- 1\,GHz IFs. Two
identical digital backends are then used to process each of these two
frequency bands. Each one consists of a Roach FPGA board and two iADC
cards \citep{2016JAI.....541001H}. The iADC cards provide dual channel sampling at
1 GHz and 8-bits resolution. The lower IF band is sampled in its
first Nyquist zone, while the upper IF band is directly sampled in the
second Nyquist zone with no further analogue downconversion. The Roach
board uses a Xilinx Virtex~5 FPGA to carry out the signal processing
tasks. The incoming signals are first channelised using a polyphase
filter bank (PFB) into 64 frequency channels of bandwidth $500/64 =
7.8125$~MHz. The PFB provides better than 40 dB of isolation between
different channels. The signals are then combined on a
channel-by-channel basis to produce the radiometer and polarimeter
outputs. A bank of complex gain corrections allows phase and amplitude
variations across the band due to the analogue part of the signal path
to be calibrated out. The sum and difference of the pairs of input
channels yields the RCP and LCP signals and their respective
reference load signals.  These are squared and averaged to provide
measures of the power in the respective sky and reference
channels. Unlike the northern system, the sky and reference signals
are not differenced in the real-time system but stored separately, and
only differenced in the off-line software. This allows us to assess
the degree of low-frequency drifts in the raw data, but which are then
cancelled out when the sky and reference are differenced. The LCP and
RCP voltage signals are complex correlated to produce the polarization
outputs $Q$ and $U$. The data are again averaged to 10~ms samples
before being read out and stored on disk by the control system.

%% file: analysis.tex
\section{Data analysis}
\label{sec:data}

\subsection{Calibration}

Accurate calibration is the key to the useful application of C-BASS data. It is essential to be able to calibrate the absolute intensity (temperature) scale, the relationship between polarized and unpolarized intensity, the absolute polarization angle, and the cross-polarization response of the instrument.

Tau A is by far the brightest polarized source that
is unresolved at C-BASS resolution, and is visible from both
observing sites. It therefore provides our primary astronomical calibration source. Observations of other bright calibrators such as Cas A are also used when Tau A is not visible (for intensity only). Observing Tau A for long continuous periods, during which the polarization angle rotates due to parallactic rotation, allows us to measure and hence correct for the non-orthogonality of the nominal $Q$ and $U$ channels. Observations of Tau A also provide the primary flux-density calibration of the data. Converting this to a temperature scale requires a knowledge of the effective area of the antenna, or equivalently of its beam pattern. We use a detailed physical model of the antenna to construct a full-sky beam pattern using the GRASP physical optics package, which is verified with comparison measurements of the main beam and sidelobes over a wide range of angles \citep{Holler2011}.   

Between primary calibration
observations, the gain and polarization angle response of the
instrument is tracked using a noise diode. A noise diode signal is
split and injected into both circular polarization channels
immediately after the linear-to-circular converter, using $-30$~dB
coaxial couplers. The diode is temperature stabilized in order to
provide a fixed-amplitude reference signal in both intensity and
polarization. The noise diode is switched on for a few seconds at the
beginning of each scan, which provides a gain measurement on a
timescale of minutes.  It provides a constant signal in both the
$I$ and $Q$ channels (in instrument co-ordinates). Phase variations
between LCP and RCP in the subsequent signal chains result in some of
the noise diode signal appearing in the instrumental $U$ channel. The
polarization data are rotated in post-processing to put the noise
diode signal wholly back into instrumental $Q$. The absolute
polarization angle will ultimately be fixed by measurements using the C-BASS South telescope of a ground-based polarized calibration source, whose polarization angle can be set to $\sim 0.1$ deg accuracy. 

Gains of both the intensity and polarization data derived from the
noise diode are interpolated to provide a continuous relative gain
correction across the entire data set. The absolute flux-density scale is set
from observations of Tau A, corrected for opacity variations between
the elevation of observation of Tau A and the elevation of the survey
scans. Since the noise diode is effectively a source of 100\% polarization (perfectly correlated between RCP and LCP), it can be used to transfer the astronomical intensity calibration to the polarized intensity calibration, so that measurements of $I$, $Q$ and $U$ are on the same scale.

The opacity is monitored by sky dip observations that are done
periodically throughout the survey observations. The telescope is
scanned between elevations 60\degr\ and 40\degr\ at a fixed
azimuth, providing a change in airmass of about 0.4, which gives a
change in background temperature of about 1.5\,K.  This signal is fitted
to a cosec(elevation) law to derive a zenith sky temperature and hence
a zenith opacity. Opacity obervations are not made below elevation 40\degr\,
to avoid contamination from ground pick-up. Opacity corrections are typically of order 1 per cent or less. 

Pointing calibration is determined from cross-scans of bright radio
sources, to which a beam model is fitted to obtain azimuth and
elevation offsets. These are then used to fit for a pointing model
incorporating collimation, axis misalignment and flexure terms. Pointing residuals are in the range of a few arcmin and are not expected to be a significant issue in data analysis.

\subsection{Flagging and data correction}

Given the relatively high temperature sensitivity of \mbox{C-BASS} (NET
$\sim 2$\,mK\,s$^{1/2}$) compared to the brightness of the sky (several
K in the Galactic plane), the C-BASS time-ordered data are frequently
signal-dominated rather than noise-dominated. This complicates the
removal of non-astronomical signals from the data. For example, it is
not possible to flag for sporadic radio-frequency interference (RFI)
simply using an amplitude clip, as a threshold low enough to eliminate
significant RFI would also flag much true emission in the sky. Instead
we use a sky model that is interpolated onto the time-ordered data
stream and subtracted. Discrepant events can then be detected and
flagged. Very small pointing errors during the crossing of bright
and/or compact sources can still generate significant residuals, so
RFI flagging is disabled for bright parts of the sky model. RFI that
is coincident with bright emission has a proportionally smaller effect
on the final map, and the very high level of redundancy in the C-BASS
observations, with each sky pixel being observed dozens of times,
means that any residual contamination is effectively washed out in the
final map. The sky model used for RFI removal is initially made using
a crude RFI cut, and progressively updated with more refined edits of
the time-ordered data.

The other main non-astronomical component of the data is ground
pick-up, which appears as a clear pattern repeating with azimuth,
and varies on timescales of many days with changes in temperature and
emissivity of the ground. As with RFI removal, a sky model is
used to subtract the bulk of the sky signal from the time-ordered data,
and regions of high sky brightness are excluded completely. The
remaining data are averaged into azimuth bins, constructing a ground
profile for every day. These profiles are then subtracted from the
data before map-making. This procedure also removes fixed RFI, such as
from fixed radio links and geostationary satellites.

\subsection{Mapping}

Although the receiver has been designed to suppress $1/f$ noise in
both intensity and polarization as much as possible, there are
long-term variations in background level, and residual atmospheric and
ground-spill emission, that are still present in the time-ordered
data. Typical $1/f$ knee frequencies in real data are around 0.1 -- 0.2 Hz. While drifts longer than a complete azimuth scan can be filtered
from the time-ordered data, shorter drifts will appear in maps as
stripes along the scan directions. However, it is possible to solve
for a good approximation to the true sky map in the presence of
drifts, using the redundancy introduced by the repeated coverage of
every pixel in the sky many times in the total time stream. Many
mapping codes have been developed to solve this problem in the context
of CMB observations \citep[e.g.,][]{Ashdown2007}, either by explicitly modelling the
drift signal or by solving the map-making equation using the full
noise statistics of the data. We use a destriping mapper, {\sc Descart}
\citep{Sutton2010}, which models the time-ordered data as consisting of
a true sky signal $s_p$ that depends on the pointing in celestial
coordinates, plus an offset series consisting of a set of constant
values $a_i$, plus stationary white noise $w_t$, i.e.,
$$
d_t = P_{tp}s_p + F_{ti}a_i + w_t.
$$$P_{tp}$ is the pointing matrix that gives the telescope
pointing direction $p$ at each time sample, and $F_{ti}$ defines the
timebase on which the offsets vary. For a well-sampled data set it is
possible to solve for the offset vector $a$, which {\sc Descart} does using
a conjugate gradient method. The offsets are then subtracted from the
data, leaving a clean time-ordered data set with only white noise,
which can be mapped by binning into sky pixels.

%% file: impact.tex
\section{Potential impact of C-BASS}
\label{sec:impact}

The C-BASS data are primarily intended to improve foreground
separation for CMB analysis by breaking degeneracies that currently exist in the component separation problem. Here we make some estimates of 
the degree of improvement in the accuracy of CMB and foreground component
parameters that can be expected from C-BASS data. 

We have simulated the component separation
process for a variety of mock data sets representing typical levels of
foreground contamination in pixels across different regions of the
sky, using the properties of existing or planned sky surveys, with and
without C-BASS.  We
assess the ability to recover a set of input parameters describing the
CMB and foregrounds, using measurements at different frequencies $\nu$ with
error bars $\sigma_{\nu}$ corresponding to particular surveys (see Table \ref{table:sensitivities} for the actual frequencies and sensitivities used). The simulations
consider only the thermal noise on a single pixel, and thus do not
include effects due to sample or cosmic variance, nor the improvement
in thermal signal-to-noise from observing a larger sky area. The full set of results showing the impact of C-BASS data on component
separation in a variety of sky regions with different levels of
foreground contamination will be presented in a forthcoming paper (Jew et al., in prep.). Here we will show representative results for one
scenario in intensity and one in polarization.

In each case, we generate mock data at each frequency for which we expect to have an observation, using a model of the foregrounds and the CMB component. We then attempt to recover the parameters from which the mock data were generated, using an MCMC fitting process. Many examples of similar techniques can be
found in the literature, including {\sc FGFIT} \citep{Eriksen2006}, {\sc Commander} \citep{Eriksen2008b}, and {\sc Miramare} \citep{Stompor2009}, and a similar methodology has been used by \citet{2018ApJ...853..127H} to explore the impact of different dust models on CMB component separation. We assign priors appropriate to the particular foreground component model. For power-law components of the form $A(\nu/\nu_0)^{\beta}$, we use the form of the Jeffreys prior $\mathcal{P}$ suggested by \cite{Eriksen2008}, namely  $\mathcal{P}(A) = 1$ and $\mathcal{P}(\beta) = [\Sigma_{\nu}(\sigma_{\nu}^{-1}(\nu/\nu_0)^{\beta}\ln(\nu/\nu_0))^2]^{1/2}$. For the CMB amplitude we use a flat prior. We also use flat priors for the amplitude and peak frequency of the AME spectrum.

We do not add noise to the mock data, so that the results are not
biased by individual realisations of the noise, but simply use the
noise levels $\sigma_{\nu}$ in the calculation of the likelihood in the fitting
process. Thus the posterior probability density functions that we show
should be interpreted as the distribution from which any particular
pixel realization would be drawn, for the given set of parameters. For example, for the intensity simulations in which we assume a CMB pixel value of $75 \, \mu \rm{K}$, the posterior density is the probability of obtaining a particular value for that pixel alone. A real observation would contain many pixels with different individual CMB values, and the CMB power would be inferred from the ensemble of pixels.

\begin{table}
\caption{The surveys and sensitivities used for the simulations. Sensitivities for the intensity simulations are for a 1\degr\ pixel while those for polarization are for a 3\degr\ pixel. The {\it FutureSat} sensitivities are taken from an early version of the LiteBIRD mission description \citep{Matsumura2014} and are intended to be indicative of a near-future satellite mission. The effective sensitivity on the Haslam map is taken to be 10 per cent of the median map temperature, i.e. it is dominated by the overall 10 per cent calibration uncertainty rather than the thermal noise.\label{table:sensitivities}}
\begin{tabular}{cccc}
\hline
Survey & Frequency / GHz & $\sigma^I \, /\mu{\rm K_{RJ}}$ & $\sigma^P \, /\mu{\rm K_{RJ}}$\\
\hline
Haslam et al & 0.408 & $2.5 \times 10^6$ &\\
\hline
C-BASS & 5.0 & 73.0 &  24.0\\
\hline
WMAP K & 22.8 & 5.8 & \\
WMAP Ka & 33.0 & 4.2 & \\
WMAP Q & 40.7 & 3.5 & \\
WMAP V & 60.7 & 3.8 & \\
WMAP W & 93.5 & 3.9 & \\
\hline
{\it Planck} 30& 28.4& 2.5& 1.1\\
{\it Planck} 44& 44.1 & 2.6&  1.3\\
{\it Planck} 70&70.4 &  3.1& 1.5\\
{\it Planck} 100&100 & 1.0& 0.51\\
{\it Planck} 143&143 &  0.33& 0.24\\
{\it Planck} 217&217 &  0.26& 0.20\\
{\it Planck} 353&353 &  0.2& 0.19\\
{\it Planck} 545&545 & 0.086 & \\
{\it Planck} 857&857 & 0.032 & \\
\hline
{\it FutureSat} 60& 60 && 0.052\\
{\it FutureSat} 78&78 & & 0.031\\
{\it FutureSat} 100&100 & & 0.020\\
{\it FutureSat} 140&140 & & 0.013\\
{\it FutureSat} 195 &195 & & 0.0070\\
{\it FutureSat} 280 &280 & & 0.0038\\
\hline
\end{tabular}
\end{table}

\subsection{Intensity}

To simulate the data we use a simplified version of the foreground
model found in Table 4 of \citet{Planck2015_X}. Our model for total
intensity measurements is summarized in Table \ref{table:intensity-model}, and consists of the following components: a single
power-law synchrotron component with amplitude $A_s$ and spectral
index $\beta_s$; a free-free component with a fixed electron
temperature of 7000\,K and effective emission measure EM; a thermal
dust component with a modified blackbody spectrum with amplitude
$A_{\rm d}$, an emissivity index $\beta_{\rm d}$ and a temperature $T_{\rm d}$; and a
single AME component with the {\sc spdust2} spectrum \citep{Ali-Hamoud2009,Silsbee2011}
allowed to shift in logarithmic frequency-brightness space with an amplitude $A_\textrm{AME}$ and peak frequency $\nu_{\rm peak}$ (following the same prescription as in \citealt{Planck2015_X}). 

\begin{table*}
\caption{The models used to generate foregrounds and CMB spectra. The free parameters are those fitted for in the MCMC fitting, while the fixed parameters are fixed for each model component and are not fitted for. Each model is used to generate a temperature component in Rayleigh-Jeans brightness temperature. \label{table:intensity-model}}
\begin{tabular}{cccc}
\hline
Component & Free parameters & Fixed parameters & Model for $T_{\rm RJ}$\\
\hline
Synchrotron & $A_{\rm s}, \beta_{\rm s}$ & $\nu_0 = 408\, {\rm MHz\,(intensity)}$& $A_{\rm s} (\nu/\nu_0)^{\beta_{s}}$\\
       &      &   $\nu_0 = 30 \,{\rm GHz\,(polarization)}$  &\\
\hline
Free-free & EM & $T_{\rm e} = 7000 \, \rm K$, $\nu_0 = 1\, {\rm GHz}$& $T_{\rm e}(1- \exp(-\tau))$\\
 & & & $\tau = 0.05468T_{\rm e}^{-3/2}\,{\rm EM}\,g_{\rm ff}\,(\nu/\nu_0)^{-2}$, \\
 & & & $g_{\rm ff} = \ln(\exp[5.96-\sqrt{3}/\pi \ln((\nu/\nu_0)(T_{\rm e}/10^4)^{-3/2})]+e)$\\
 \hline
 AME & $A_{\rm AME}, \nu_{\rm peak}$ & & {\sc spdust2} \\
\hline
Dust & $A_{\rm d}, \beta_{\rm d}, T_{\rm d}$& $\nu_0 = 545\, {\rm GHz\,(intensity)}$ &$A_{\rm d}\Big(\frac{\nu}{\nu_0}\Big)^{\beta_{\rm d}+1}\, \frac{\exp(h\nu_0/k_{\rm B} T_{\rm d})-1}{\exp(h\nu/k_{\rm B} T_{\rm d})-1}$ \\
 & & $\nu_0 = 353 \, {\rm GHz\,(polarization)}$& \\
\hline
CMB & $A_{\rm CMB}$& $T_0 = 2.7255 \, {\rm K}$ & $A_{\rm CMB}\, x^2 e^x / (e^x - 1)^2$,\\
& & & $x = h\nu / k_{\rm B}T_0$\\
\hline
\end{tabular}
\end{table*}

\begin{table*}
\caption{Recovered parameter values for the intensity simulations, with and without the inclusion of the C-BASS data point (corresponding to the posterior density estimates in Fig. \ref{fig:sim_GP_I_pdf}. }
\begin{tabular}{lrrrr}
\hline
Parameter  &   Recovered value &  Recovered value & True value & Units\\
           &    (No C-BASS)    & (with C-BASS)    &\\

\hline
$A_\textrm{s}$ @ 100\,GHz & $1.33_{-1.33}^{+1.81}$ & $1.84_{-0.165}^{+0.191}$ & $1.86$ & $\mu$K$_\textrm{RJ}$ \\
$\beta_\textrm{s}$ & $-3.02_{-0.16}^{+0.11}$ & $-3.10_{-0.026}^{+0.025}$ & $-3.10$ &  \\
EM & $365_{-21}^{+11}$ & $362_{-4}^{+4}$ & $361$ & cm$^{-6}$pc \\
$A_\textrm{AME}$ & $701_{-39}^{+37}$ & $707_{-11}^{+13}$ & $708$ & $\mu$K$_\textrm{RJ}$ \\
$\nu_\textrm{peak}$ & $25.0_{-3.2}^{+3.1}$ & $25.0_{-1.6}^{+1.4}$ & $25.0$ & GHz \\
$A_\textrm{d}$ & $2080.9_{-0.11}^{+0.10}$ & $2080.9_{-0.09}^{+0.09}$ & $2080.86$ & $\mu$K$_\textrm{RJ}$ \\
$\beta_\textrm{d}$ & $1.545_{-0.00087}^{+0.00095}$ & $1.545_{-0.00074}^{+0.00097}$ & $1.545$ &  \\
$T_\textrm{d}$ & $17.480_{-0.012}^{+0.011}$ & $17.481_{-0.012}^{+0.009}$ & $17.480$ & K \\ 
$A_\textrm{CMB}$ & $75.4_{-2.3}^{+2.0}$ & $75.0_{-1.2}^{+1.3}$ & $75.0$ & $\mu$K$_\textrm{CMB}$ \\
\hline
\end{tabular}
\label{table:results_I}
\end{table*}

\begin{table*}
\caption{Recovered parameter values for the polarization simulations, with and without the inclusion of the C-BASS data point, corresponding to the posterior density estimates in Fig.  \ref{fig:sim_B_pdf}.}
\begin{tabular}{lrrrr}
\hline
Parameter  &   Recovered value &  Recovered value & True value & Units\\
           &    (No C-BASS)    & (with C-BASS)    &\\
\hline
$A_\textrm{s}$ @ 100\,GHz & $0.086_{-0.048}^{+0.149}$ & $0.072_{-0.018}^{+0.021}$ & $0.074$ & $\mu$K$_\textrm{RJ}$ \\
$\beta_\textrm{s}$ & $-2.37_{-0.27}^{+1.37}$ & $-3.09_{-0.10}^{+0.08}$ & $-3.10$ &  \\
$A_\textrm{d}$ & $0.313_{-0.023}^{+0.034}$ & $0.329_{-0.019}^{+0.022}$ & $0.335$ & $\mu$K$_\textrm{RJ}$ \\
$\beta_\textrm{d}$ & $0.97_{-0.96}^{+0.37}$ & $1.56_{-0.50}^{+0.51}$ & $1.63$ &  \\
$T_\textrm{d}$ & $65.8_{-36.6}^{+4.2}$ & $65.3_{-34.9}^{+4.7}$ & $24.9$ & K \\ 
$A_\textrm{CMB}$ & $-0.02_{-0.38}^{+0.09}$ & $0.02_{-0.09}^{+0.06}$ & $0.00$ & $\mu$K$_\textrm{CMB}$ \\
\hline
\end{tabular}
\label{table:results_P}
\end{table*}

We use the component separation
results from \citet{Planck2015_X} to suggest values of the foreground
parameters. For this example, we used a region close to the Galactic plane to illustrate a fairly severe instance of foreground contamination. We then produce mock brightness values using the foreground models plus a CMB signal.  
We simulate the intensity measurements
in 1\degr\ pixels, since all components (including the CMB) are
detected at high signal-to-noise ratio in a typical pixel. The CMB value was set to 75\,$\mu$K, corresponding to the rms fluctuations on a 1\degr\ scale. Simulated observations at the central
frequencies of the Haslam, {\it Planck}, WMAP and C-BASS surveys were
included. For each
frequency measurement we assigned thermal noise based on the
achieved or expected sensitivity of the appropriate survey.  These are summarized in Table \ref{table:sensitivities}.

\begin{figure*}
  \centering
  \includegraphics[width=1.0\textwidth]{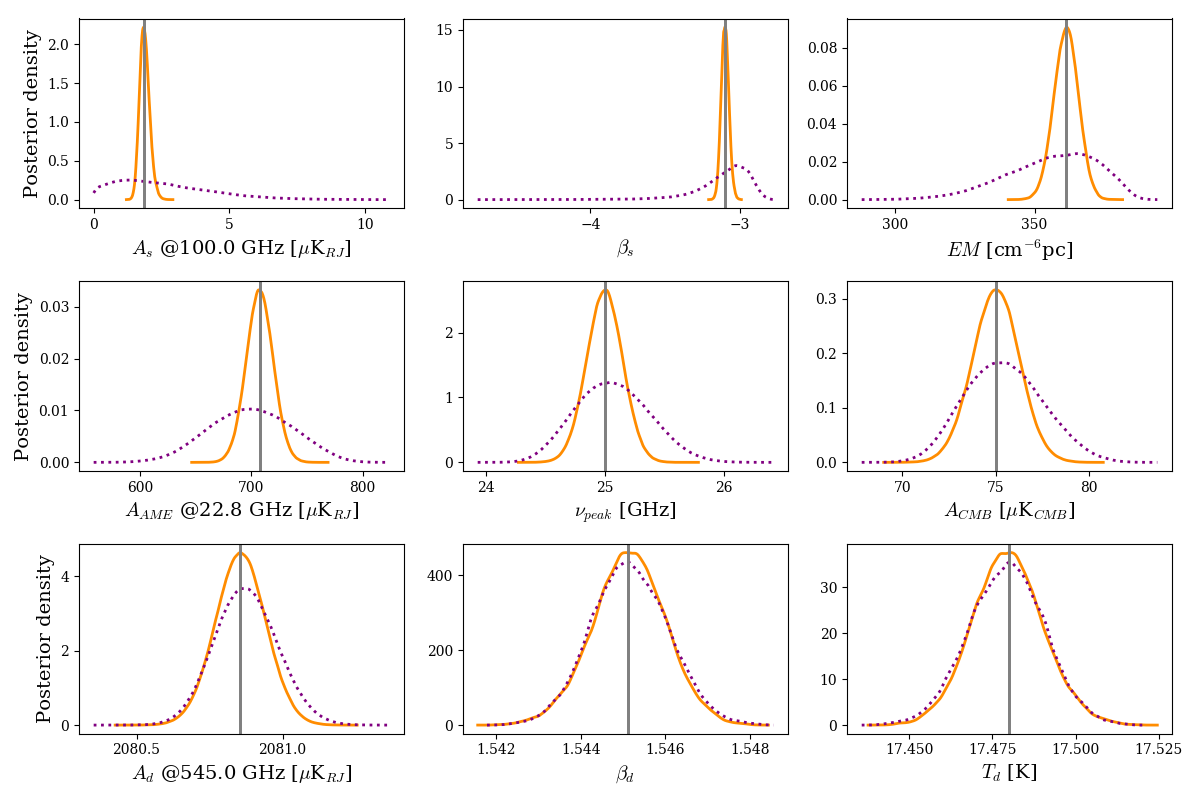}
  \caption{PDEs of the total intensity component parameters
           for a typical 1\degr\ pixel in a sky region with significant foreground contamination. 
           The dashed lines are the PDEs when only including 
           Haslam, WMAP and {\it Planck} data points in the fit. The solid lines are the PDEs 
           when the C-BASS data point is included. The vertical lines are
           at the true parameter values used to simulate the data.}
  \label{fig:sim_GP_I_pdf}
\end{figure*}

\begin{figure*}
  \centering
  \includegraphics[width=1.0\columnwidth]{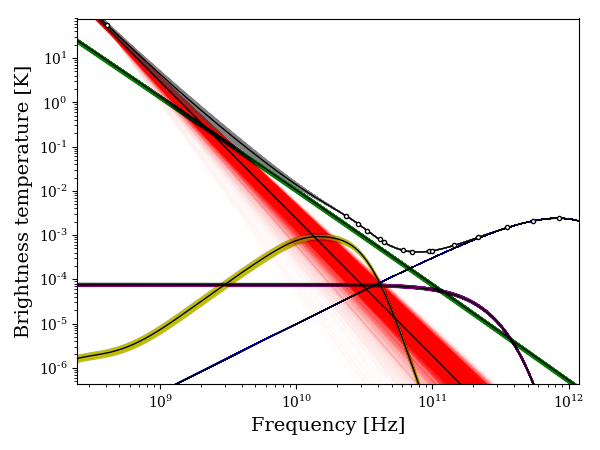}
  \includegraphics[width=1.0\columnwidth]{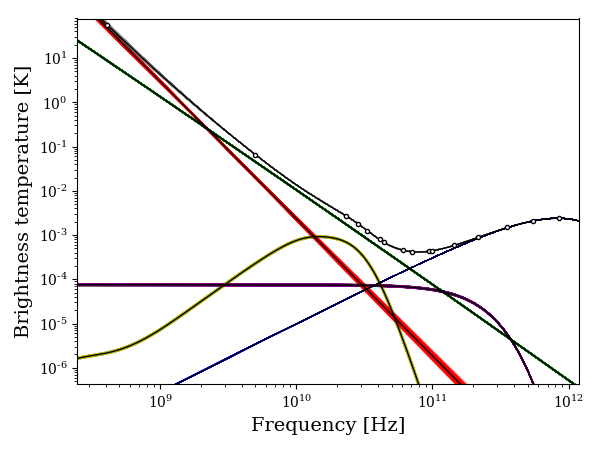}
  \caption{Total intensity frequency spectra for a 1\degr\ pixel in a sky region with significant foreground contamination.
           The solid black lines are spectra of the true simulated foreground components.
           The coloured lines are the frequency spectra of the sky components of 5000 
           randomly drawn samples from the converged MCMC chains. {\it Left} is the result from 
           only including Haslam, WMAP and {\it Planck} data points. {\it Right} is with the addition of 
           a C-BASS data point. Synchrotron is { red}; thermal dust is {blue}; AME is {yellow}; 
           free-free is {green}; and CMB is { purple}.}
  \label{fig:sim_GP_I_spectrum}
\end{figure*}

Figure~\ref{fig:sim_GP_I_pdf} shows the posterior density estimates
(PDE) of the total intensity foreground parameters for a single 1\degr\
pixel in a region with significant AME and free-free
emission. Figure~\ref{fig:sim_GP_I_spectrum} shows the corresponding
estimates of the actual component spectra, along with the true input
spectra, and Table \ref{table:results_I} shows the numerical values for the recovered parameters. These are given as the peak posterior value and the parameter range that contains 68 per cent of the posterior volume, as the PDEs are often quite skewed and cannot be represented with a symmetrical error bar. Without the C-BASS data, the synchrotron parameters, $A_{\rm s}$ and $\beta_s$, are very poorly constrained. Including the C-BASS data improves the measurement of the synchrotron radiation amplitude by an order of magnitude, and reduces the error range on the spectral index from 0.27 dex to 0.05 dex. It also markedly improves the estimates of the free-free emission measure and the AME parameters, reducing the error bars on these parameters by factors of 2 -- 4. There is even a small improvement on the constraints on the dust amplitude. These improvements in foreground parameter estimates result in a reduction of the errors on the measurement of the CMB amplitude in this pixel of 40 per cent. 

\subsection{Polarization}

For the polarization simulations we did not include a free-free or AME
component. Free-free emission is essentially unpolarized, while AME polarization is expected to be small, and has not yet been detected. We also set the CMB signal to zero. This represents a situation in which the $E$-mode signal has been perfectly separated out, and we are searching for a $B$-mode signal of very small amplitude. Data points at the centre frequencies of C-BASS and {\it Planck} are included, along with a set of sensitivities indicative of a near-future CMB satellite mission (`{\it FutureSat}'), based on the early mission description of {\it LiteBIRD} \citep{Matsumura2014}.

\begin{figure*}
  \centering
  \includegraphics[width=1.0\textwidth]{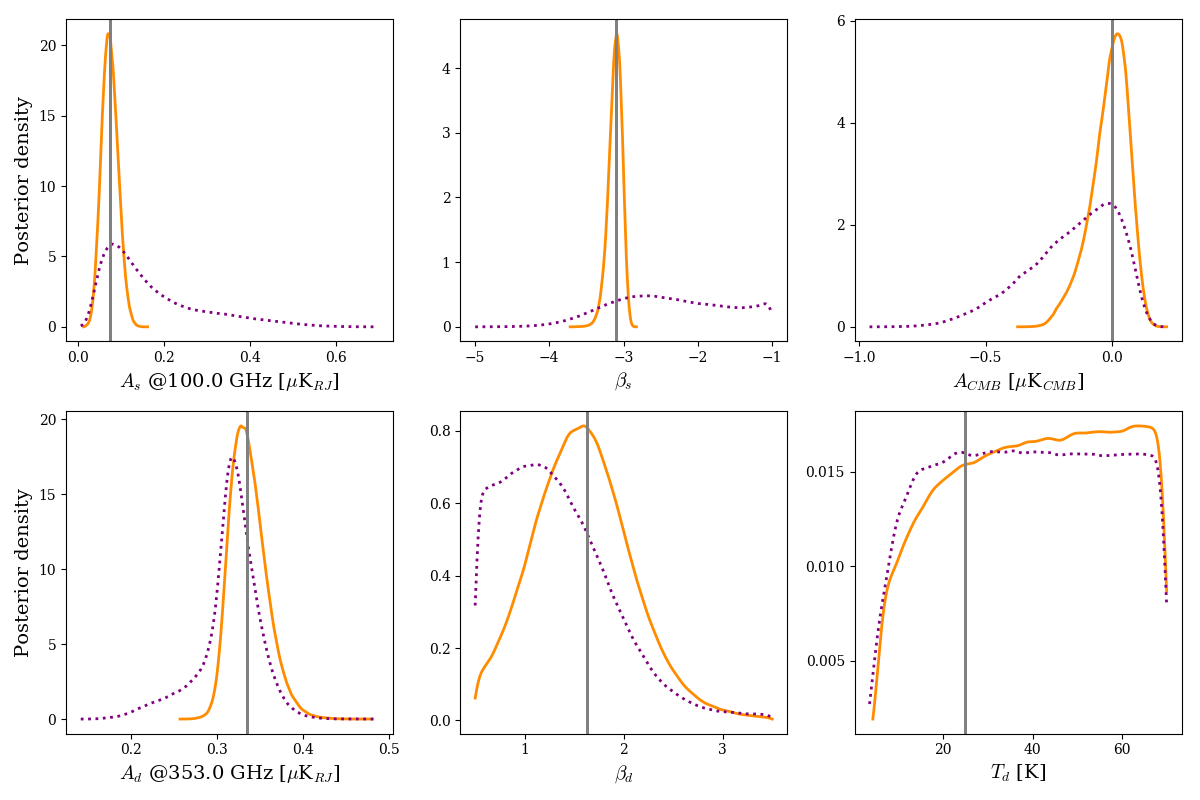}
  \caption{PDE of the $B$-mode polarization component parameters for a typical 
           3\degr\ pixel in a sky region with low foreground emission.
           The dashed lines are the PDEs when only including
           {\it Planck} and {\it FutureSat} data points. The {solid lines} are the posterior
           density estimates when the C-BASS data point is included. The vertical lines are
           at the true parameter values used to simulate the data.}
  \label{fig:sim_B_pdf}
\end{figure*}

\begin{figure*}
  \centering
 \includegraphics[width=1.0\columnwidth]{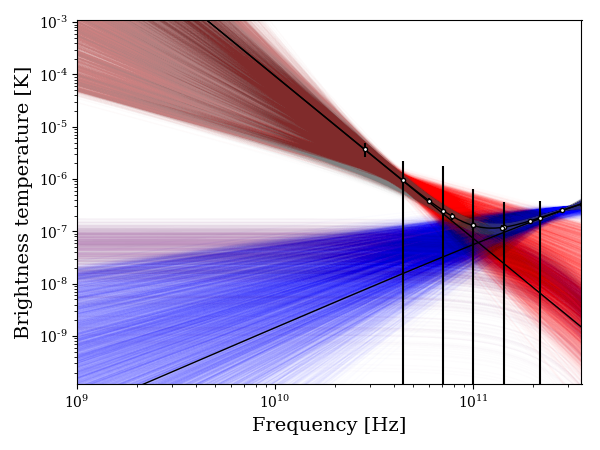}
  \includegraphics[width=1.0\columnwidth]{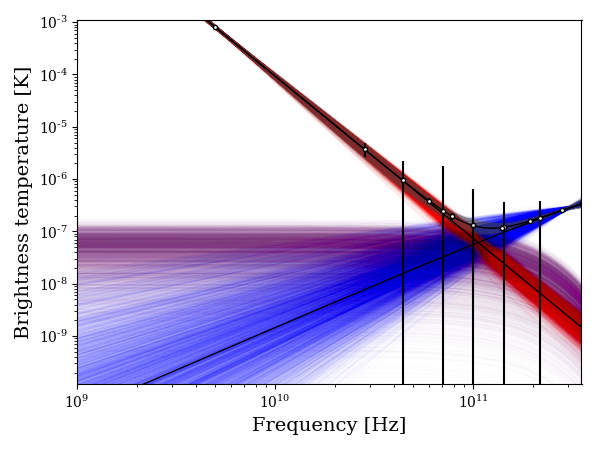}
  \caption{B-mode polarization frequency spectra for a 3\degr\ pixel in a sky region with low foreground            emission. The {solid black lines} are spectra of the true simulated foreground components.
           The coloured lines are the frequency spectra of the sky components of 5000 
           randomly drawn samples from the converged MCMC chains. {\it Left} is the result from 
           only including {\it Planck} and {\it FutureSat} data points. {\it Right} is with the addition of 
           the C-BASS data point. Synchrotron is { red}; thermal dust is { blue}; and CMB is { purple}.}
  \label{fig:sim_B_spectrum}
\end{figure*}

The PDE of the polarization foreground parameters ($B$-mode) 
for a 3\degr\ pixel in a
low-foreground region of sky are shown in
Figure~\ref{fig:sim_B_pdf}. Figure~\ref{fig:sim_B_spectrum}
shows the corresponding estimates of the component spectra, along with
the true input spectra, and Table \ref{table:results_P} summarizes the results.  
Including C-BASS data results in much tighter constraints on the synchrotron amplitude and spectral index, with a previously almost unconstrained spectral index now measured with an accuracy of 0.1 dex. There is also significant improvement in the dust spectral index, resulting in a reduction in the 1-$\sigma$ range on the CMB amplitude by a factor of three. Additional
low-frequency points between the C-BASS and {\it Planck} frequencies would
provide additional constraints on the synchrotron spectrum and lower
bias on the $B$-mode amplitude measurement. 

While the addition of the C-BASS data point dramatically improves the recovery of the synchrotron components and the CMB amplitude in the case of a straight synchrotron spectrum, additional complication in the synchrotron spectra will require additional observational constraints. A C-BASS-like instrument covering frequencies between 5~GHz and the lower end of the space microwave band would provide constraints on realistic synchrotron spectra, including the effects of intrinsic curvature and line-of-sight integration of different spectra. A detailed study of such an instrument, NextBASS, and its potential impact on component separation using the techniques presented here, is in preparation.

